\documentclass[12pt]{article}
\usepackage{graphicx}
\begin{document}

\title{Isotopic trends in nuclear multifragmentation}

\author{Martin Veselsky\\
\\
Institute of Physics, Slovak Academy of Sciences,\\
Dubravska cesta 9, Bratislava, Slovakia\\
e-mail: fyzimarv@savba.sk
}

\date{}

\maketitle

\begin{abstract}
An overview of the recent progress in the studies of nuclear 
multifragmentation is presented.
Special emphasis is put on the exploration of isotopic trends in nuclear
multifragmentation and the possibilities to extract physical information
related to the nuclear equation of state. 
Relevant experimental methods of isotope identification are described. 
The isotopic composition of fragments 
is used to extract the values of thermodynamical observables of the system
undergoing multifragmentation such as temperature and chemical potentials. 
Various methods for extraction of thermodynamical variables are analyzed.
An overview of methods of isotope thermometry, exploring the sensitivity of 
various yield ratios to temperature, is presented. 
An exponential scaling of relative isotopic yields from 
reactions with different neutron content, called isoscaling, 
is used to explore the evolution of the isospin degrees of freedom 
of the system.  Finally, the nuclear equation of state and 
the isospin-asymmetric liquid-gas phase transition in the nuclear matter 
are discussed.
\end{abstract}

\section*{Introduction}

Experimental studies, carried out during last two decades 
( for review see \cite{ExpMfrag} and for update of the recent progress 
see e.g. \cite{ExpMfragRec} ), demonstrate 
that, in nuclear reactions such as the nucleus-nucleus collisions in the Fermi 
energy domain or the collisions of a high-energy light particle with a heavy 
target, the amount of kinetic energy dissipated into the internal excitation 
energy is large enough to produce highly excited ( hot )  nuclei which may 
further de-excite via multifragmentation, a decay mode where the hot nucleus 
disintegrates 
into many fragments. In the nature, nuclear processes of such type can be 
observed in the interaction of cosmic rays with terrestrial and interplanetary 
matter. Such spallation reactions contribute to the observed isotopic 
composition of the interstellar matter which can be used to study 
the properties of the cosmic rays and of their astrophysical sources 
( for review see e.g. \cite{AstroSpall} ). With the advent of powerful 
accelerators of high-energy particles and, more recently, of heavy ions 
of intermediate and high energy, it became possible to study the process 
of multifragmentation in the laboratory. Extensive experimental 
studies of multifragmentation are conducted since eighties of the last century. 
As can be deduced from observed experimental data \cite{ExpMfrag,ExpMfragRec}, 
during the process 
of multifragmentation the nuclear medium becomes hot,  expands and 
is supposed to undergo a phase transition where two phases 
are created, a dense isospin-symmetric phase ( liquid ) and a dilute 
isospin-asymmetric phase ( vapor or gas ). Such a phase transition was 
predicted by theory ( for review see \cite{Richert} ). 
The most common theoretical 
description of multifragmentation is based on the statistical model  
of multifragmentation \cite{SMM}, typically assuming a thermally equilibrated 
freeze-out configuration with hot ( excited ) or cold fragments. 
The probabilities of various fragment partitions can be evaluated 
analytically using the grand-canonical approach. While the statistical 
model of multifragmentation is quite successful in describing a wide range of 
fragment observables, it does not provide a specific information on the 
mechanism of the underlying process of fragment formation, apart from the 
assumption that at the freeze-out configuration a thermal equilibrium among 
the fragments is reached. The underlying process of cluster formation 
can be related to the equation of state of asymmetric nuclear matter 
at sub-saturation densities, where the phase transition is supposed to occur. 
The coexistence line, 
typical for the one-component system such as symmetric nuclear matter, 
develops with introduction of isospin asymmetry into a complex coexistence 
region \cite{MullerSerot}, typically suggesting a coexistence of the dilute 
isospin-asymmetric
phase and isospin-symmetric dense phase. Further 
theoretical investigations \cite{Spinod} suggest that instability modes 
responsible 
for the phase transition are typically a combinations of mechanical and 
chemical instability modes.

In this article, we present an overview of the recent progress in the studies 
of nuclear multifragmentation.
We place a special emphasis on the exploration of isotopic trends in nuclear
multifragmentation and the possibilities to extract physical information
related to the nuclear equation of state. 
Experimental studies of isotopic trends 
in nuclear multifragmentation have been made possible by use of the 
state-of-the-art detection techniques, which allow to identify 
individual isotopes of light charged particles and intermediate mass fragments 
in wide angular ranges. The isotopic composition of fragments can be used
to extract the values of thermodynamical observables of the system
undergoing multifragmentation such as temperature and 
chemical potentials ( free nucleon densities ).
Various methods for extraction of temperature from isotopic composition 
are analyzed. 
An exponential scaling of the yields of specific isotopes from different
reactions with the isotope neutron and proton numbers ( isoscaling ) 
is used to study 
the evolution of the isospin degrees of freedom of the system. Finally, the
nuclear equation of state and the isospin-asymmetric liquid-gas phase
transition in the nuclear matter are discussed.

\begin{figure}[h]
\centering
\vspace{5mm}
\includegraphics[width=12.5cm,height=6.cm]{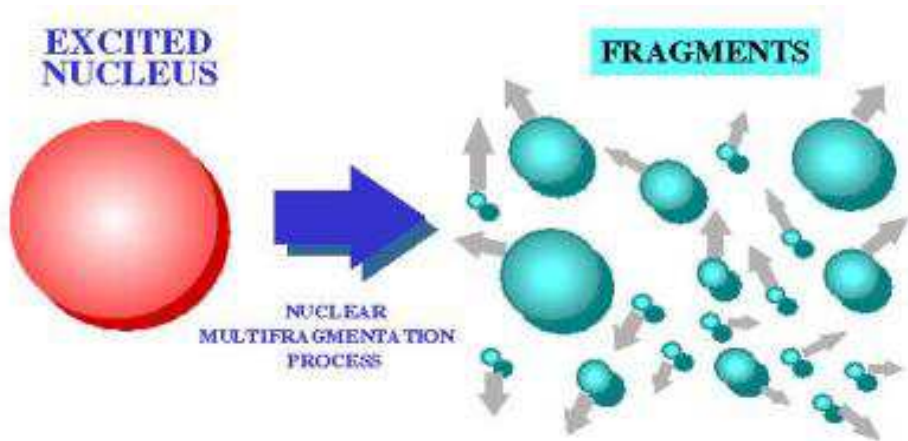}
\caption{\footnotesize  
A schematic view of multifragmentation.
}
\label{fmfrg}
\end{figure}

\section*{Multifragmentation theory and related topics}

Multifragmentation is a process where the final state consists of multiple 
fragments which are the remnants of a hot nucleus. The aim of the 
multifragmentation theory is to relate the properties of final fragments to 
properties of the original hot nucleus. Statistical phase-space models of 
multifragmentation typically assume that the properties of final fragments 
are determined in the so-called "freeze-out" configuration, where the typical 
distance of pre-fragments becomes, due to thermal expansion, larger than 
radius of the short-range 
nuclear interaction. It is, however, difficult to study the properties of 
the freeze-out configuration using the experimentally observed final 
fragments without additional model assumptions reflecting the reaction 
mechanism by which the hot nucleus was created. A detailed knowledge  
of the reaction mechanisms contributing to production of hot nuclei is 
necessary to properly characterize initial conditions of multifragmentation. 
The production of hot nuclei can be treated separately as a stage prior to 
multifragmentation, assuming that statistical equilibrium 
will be reached at later stage ( typically at freeze-out ). 
In a microscopic description of nuclear medium, multifragmentation can be 
related to a phase transition into dense and dilute phase. Isospin 
asymmetry of two phases may be different. Furthermore, multifragmentation 
can be treated dynamically using the transport theory. Such treatment 
offers the possibility to study 
the evolution of the nuclear medium during the whole process, including 
the conditions for the isospin-asymmetric phase transition. 
In this section we briefly review the most relevant aspects of 
the multifragmentation theory and related reaction theory.  

\subsection*{Statistical phase-space models of multifragmentation}

A main assumption of the statistical phase-space models describing various 
processes is that a certain stage of the process can be identified where the 
available phase-space determines the 
final probability distribution of exit channels. For instance, in 
statistical description of the fully equilibrated compound 
nucleus, the probabilities of various evaporation channels are 
determined by available phase space of the corresponding evaporation 
residue and the fission probability is determined by available phase space 
at the saddle-point on top of the fission barrier. 
In the case of simultaneous multifragmentation, the phase-space 
is explored for the possible multi-fragment 
configurations ( partitions ), typically at the stage where short-range nuclear 
interaction  freezes out and the identities of final fragments are determined. 
This is the so-called "freeze-out" configuration. The probabilities of 
various partitions can be determined statistically, exactly using 
the micro-canonical ensemble \cite{MMMC}, or approximately using 
the canonical and 
grand-canonical ensembles \cite{SMM}. Canonical and grand-canonical 
approximations 
typically reduce the mathematical complexity while raising questions on the 
applicability of such approximations since the nuclei are rather small 
systems. Detailed discussion of the application 
of statistical mechanics to small systems can be found in \cite{FinThDyn}. 

Here we briefly review the most commonly used phase-space model of 
multifragmentation, the Statistical Model of Multifragmentation ( SMM )
\cite{SMM}. 
The model uses the grand-canonical approximation, the freeze-out volume 
depends on fragment multiplicity and internal excitation 
of fragments in the freeze-out configuration is considered, thus 
producing "hot" fragments. Analogous models with "cold" fragments also 
exist \cite{MMMC}, typically the freeze-out volume is chosen approximately 
twice 
larger than in the case of hot fragments. Yield of fragments with neutron 
and proton numbers N and Z can be, within grand-canonical approach, 
characterized by formula \cite{SMM}:

\begin{equation}
        Y(N,Z) = g_{NZ} \frac{V_f}{\lambda_T} A^{3/2}
	\exp(-(1/T)(F_{NZ}(T,V)-N \mu_{n} - Z \mu_{p}))
\label{eqnynz}
\end{equation}

where $g_{NZ}$ is the ground state spin degeneracy, $V_f$ is the free 
volume ( $V_f = V - V_0$ where $V$ and $V_0$ are the volumes at freeze-out and 
ground state ), $\lambda_T$ is the thermal wavelength, $F_{NZ}$ is the 
intrinsic free energy of the fragment, $\mu_{n}$ and $\mu_{p}$ are 
the free neutron and proton chemical 
potentials, and $T$ is the temperature. The free volume at freeze-out 
is determined as $V_f = \chi V_0$. The parameter $\chi$ is determined as 

\begin{equation}
        \chi = ( 1 + \frac{d}{r_0 A_0^{1/3}} (M^{1/3}-1))^3 - 1 .
\label{eqnchi}
\end{equation}

where $M$ is the fragment multiplicity, $d$ is nuclear diffuseness, 
$A_0$ is the mass of the system and $r_0$ is the nuclear radius parameter. 
It is essentially a product of the additional fragment surface area 
and nuclear diffuseness length. The values of $\chi$ range from 0.2 to 2. 
The fragment free energy is at most part determined using the liquid drop 
model. The component of free energy corresponding to Coulomb interaction 
is determined using the Wigner-Seitz approximation. 
The Wigner-Seitz 
clusterization energy is calculated at volume $V = ( 1 + \kappa ) V_0$. 
The value of $\kappa$ can be related to average Coulomb barrier and thus 
is generally different from $\chi$, which controls the freeze-out of strong 
interaction. Typical value is $\kappa = 3.5$ which is 
between the values for closely packed fragments ( $\kappa = 2$ ) and 
for fission ( $\kappa = 5$ ).  The statistical model of multifragmentation 
was used for extensive comparisons with experimental 
multifragmentation data, typically obtaining a consistent description 
of a wide range of experimental observables \cite{SMM,SiSnNExch,SnAlMARS}.  

As an alternative to models of simultaneous multifragmentation, the 
traditional model of compound nucleus decay based on Hauser-Feshbach 
approximation \cite{HausFesch} was extended to describe the emission 
of intermediate mass fragments ( IMF ) \cite{AsymFiss}. 
The emission of such complex fragments is described as a binary split, 
essentially an asymmetric fission where the IMF is accompanied 
by a heavy residue.  The emission probability 
is determined by a value of the mass-asymmetric fission barrier height. 
The model of fission-like binary split describes reasonably  well 
the onset of fragment emission, thus indicating that at the onset 
the complex fragment emission is a fission-like process \cite{McMahan}. 
Success of such approach prompted the attempts to describe 
multifragmentation using the model of Sequential Binary Decay ( SBD ). 
Several implementations ( e.g. the GEMINI code \cite{GEMINI} ) have 
been used extensively for comparisons 
with experimental multifragmentation data, with reasonable success in 
description of e.g. inclusive mass and charge distributions. However, 
the model 
of sequential binary decay fails to describe the results of the 
exclusive measurements at the zero angle \cite{SnAlMARS} allowing to select 
a narrow region around the multifragmentation threshold 
while SMM performs rather well. 
An explanation is suggested by comparison of  
the model of sequential binary  decay to transitional-state model of 
multifragmentation \cite{MFrgTrans} where even for binary exit channel 
the number of available degrees of freedom is different. 
Furthermore, the model of sequential binary decay also fails to reproduce 
the observed fragment correlations \cite{SBDFail}.  

\subsection*{Isospin-asymmetric liquid-gas phase transition}

Phase-space models of multifragmentation typically explore the properties 
of final fragments in the freeze-out configuration where nuclear 
interaction between the fragments does not play a role. The underlying 
mechanism leading to multifragmentation is nevertheless closely related 
to properties of the nuclear medium, which can be studied 
microscopically. Over the recent decades, thermal properties of the 
nuclear medium were studied theoretically using microscopic approaches 
such as temperature-dependent Hartree-Fock \cite{TdepHF} or relativistic 
mean-field \cite{MullerSerot} 
approximations. The microscopic theory of the nuclear matter predicts a 
phase transition of the liquid-gas type. 
Below a critical temperature, the two phases can coexist. 
The spatial distribution of phases can be expected to develop 
from vapor bubbles in liquid at high density toward liquid droplets 
in vapor at low densities \cite{DropBubbl}. A hot nucleus undergoing 
multifragmentation 
can be envisioned as rapidly expanding until it enters the spinodal region. 
At that stage multifragmentation occurs due to rapidly developing dynamical 
instabilities. When considering the isospin-asymmetric nuclear matter
\cite{MullerSerot,KolSanzh}, the phase transition is supposed to lead to 
separation into a symmetric dense phase and asymmetric dilute phase. 
It has been discussed in literature
\cite{Spinod} to what extent such a phase transition
is generated by fluctuations of density or concentration, typically
suggesting a coupling of different instability modes. 

\begin{figure}[h]
\centering
\vspace{5mm}
\includegraphics[width=6.cm,height=6.cm]{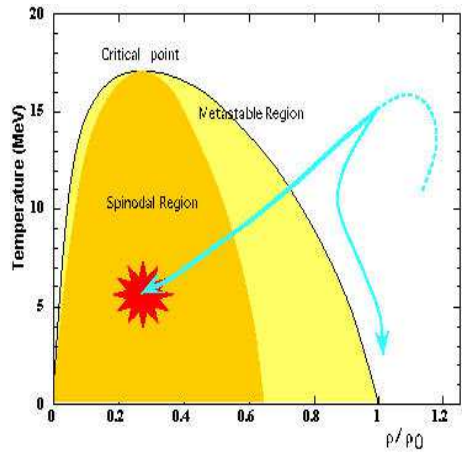}
\caption{\footnotesize  
A schematic description of the nuclear liquid-gas phase transition 
as a mechanism of multifragmentation.
}
\label{flgphtr}
\end{figure}

The properties of clusters formed in the phase transition can be 
described by a variety of random cluster models. Simple model of 
clusterization was proposed by Fisher \cite{Fisher}, where the 
liquid 
drop expansion of the Gibbs free energy leads to the mass distribution 
of clusters 

\begin{equation}
        P(A) \propto A^{-\tau} \exp(-b(T) A^{2/3})
\label{eqnpa}
\end{equation}

where the surface term $b(T)=4\pi r_0^2\sigma(T)/T$ decreases monotonically 
toward zero as temperature approaches critical temperature $T_c$. At critical 
temperature the mass distribution becomes a power law with critical exponent 
$\tau$. Similar properties are obtained using percolation models, 
lattice models and cellular model of fragmentation ( an extensive overview 
of these models in the context of multifragmentation can be found 
in \cite{Richert}, sections 4.,5. ). 

Calculations based on the microscopic nuclear theory \cite{MullerSerot,KolSanzh} 
suggest that nuclear 
matter undergoes liquid-gas phase transition which is of the first 
order, the density difference between phases being the order parameter. 
Similar conclusions can be obtained also in the context of SMM \cite{SMMOrder}. 
However, 
the phase transition in the percolation models is of the second order and 
lattice models such as Ising model \cite{Ising} exhibit a complex phase 
diagram with both first and second order phase transitions. Furthermore, since  
theory of nuclear phase transitions is formulated for infinite systems while 
the experiments investigate multifragmentation of finite nuclei, 
it is necessary to assess the effect of finite size. 
As discussed in \cite{Richert}, the critical behavior 
observed in random-cluster models can be related 
to scaling properties and self-similarity of the system at different scales, 
what suggests that observed properties of finite system represent 
the properties of infinite media. On the other hand, finiteness constraints 
of small systems can lead to a behavior analogous to criticality \cite{Richert}. 
Obviously, for any experimental observation the effect of finite size should 
be assessed.

\subsection*{Production of hot nuclei}

The statistical multifragmentation theory typically does not 
explicitly consider the evolution of the system prior to the freeze-out 
( or underlying phase transition ). Usually it is difficult to determine 
the properties of the freeze-out configuration using the experimentally 
observed final fragments without additional model assumptions reflecting 
the reaction mechanism by which the hot nucleus was created. A detailed 
knowledge  of the reaction mechanisms contributing to production of hot nuclei 
is necessary to properly characterize initial conditions of multifragmentation 
and to verify the model assumptions used by multifragmentation theory. 

In order to undergo multifragmentation, intrinsic excitation energy 
of the hot massive nuclei should exceed 2 - 3 AMeV. Such excitation energies 
can be reached in nucleus-nucleus collisions with the nuclear beams 
of intermediate energies ( 20 - 100 AMeV ). In collisions of a high-energy 
light particle with a heavy target the necessary energies range from 
hundreds of AMeV to AGeV. These are the reaction domains which have been 
used extensively for multifragmentation studies.  

\begin{figure}[h]
\centering
\vspace{5mm}
\includegraphics[width=6.cm,height=6.cm]{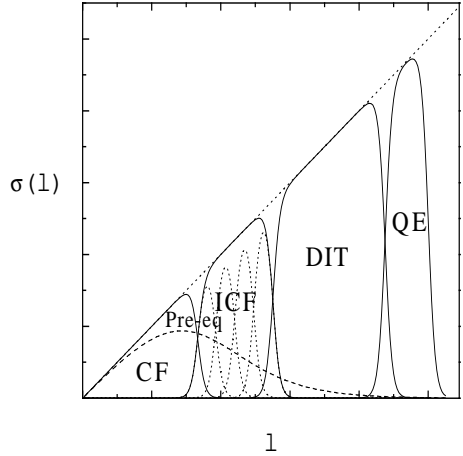}
\caption{\footnotesize  
Development of the reaction mechanisms in 
nucleus-nucleus collisions in the Fermi energy domain. 
Cross section contributions are plotted schematically as a function 
of initial angular momentum. 
}
\label{frmlsyst}
\end{figure}

A large variety of reaction mechanisms has been observed 
in nucleus-nucleus collisions in the Fermi energy domain ( 20 - 50 AMeV ) 
depending on the impact 
parameter, projectile-target asymmetry and the projectile energy. The reaction 
mechanisms typically observed are \cite{Prod}: \\

- peripheral elastic and quasi-elastic ( QE ) scattering/transfer reactions 
around the grazing impact parameter.

- deep inelastic transfer ( DIT ) reactions at semi-peripheral impact 
parameters with partial overlap of the projectile and target and a significant 
part of the relative kinetic energy transferred into internal excitation 
energies of the projectile and target.

- incomplete fusion ( ICF ) reactions at central impact parameters with 
a typical participant-spectator scenario. At energies around the Fermi energy 
the participant zone typically fuses with one of the spectator zones 
(typically heavier) thus creating a highly excited composite nucleus. At 
most central impact parameters the complete fusion ( CF ) can occur.

- pre-equilibrium emission of direct particles, caused by onset of two-body
nucleon-nucleon collisions at central and mid-peripheral impact parameters. 
Pre-equilibrium emission typically precedes the ICF/CF and DIT reactions.
\\

While such a scheme is rather well justified by experimental data \cite{Prod}, 
there 
are still many open questions related to the reaction mechanism, specifically 
concerning the development of isospin-asymmetry in nucleus-nucleus 
collisions in the Fermi energy domain. Studies of isospin degrees of freedom 
in the nucleus-nucleus collisions in the Fermi energy domain reveal many 
interesting details of the reaction scenario. Recently, an enhancement in the 
production of neutron-rich nuclei was observed in peripheral 
nucleus-nucleus collisions \cite{NSkin}, which can be related 
to the effect of neutron-rich surface of the target.

At projectile energies of hundreds of AMeV and above the impinging light 
particle initiates an intra-nuclear cascade \cite{INC} of nucleon-nucleon 
collisions 
which is the main mechanism leading to production of hot nuclei. 
The heavy remnant after the intra-nuclear cascade 
is highly excited and typically undergoes multifragmentation at wide range 
of impact parameters \cite{SMM}. In the nucleus-nucleus collisions 
at similar 
energies the participant zone develops independently of two spectator zones 
and typically disintegrates due to intense cascade of nucleon-nucleon 
collisions, with possibility to produce sub-nuclear particles via excitation 
of hadronic degrees of freedom. 
The spectator zones are relatively colder, nevertheless typically 
hot enough to undergo multifragmentation \cite{SpecMFrg}.

\subsection*{Dynamical description of multifragmentation}

As an alternative to two-step reaction scenario, where reaction dynamics 
is attributed to the early stage of nucleus-nucleus collision while 
multifragmentation of the hot nucleus is described statistically, 
the whole process can be described in unified fashion by dynamical theory 
with explicit time-dependence. 

In the microscopic time-dependent Hartree-Fock theory \cite{TDHF}, 
an extension of the Hartree-Fock theory of the ground state, 
the nucleus-nucleus 
collision is represented by time evolution of the one-body mean field, which 
defines the collective degrees of freedom of the system. 
The time-dependent Hartree-Fock theory does exhibit collective behavior 
which reminds many experimentally observed phenomena \cite{TDHF}, 
it nevertheless does 
not include two-body dissipation via collisions of nucleons 
which is supposed to play significant role 
at energy regimes where the hot nuclei are produced. 

\begin{figure}[h]
\centering
\vspace{5mm}
\includegraphics[width=7.cm,height=6.cm]{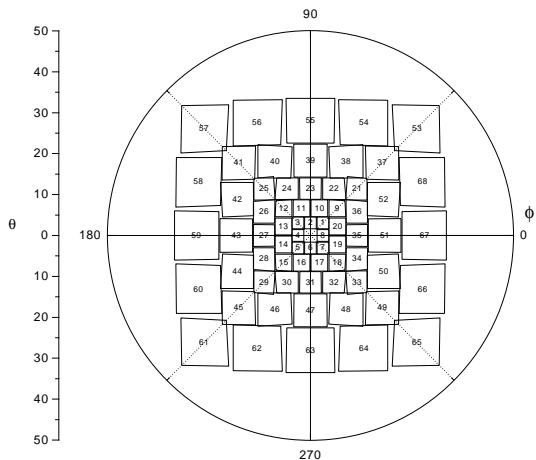}
\caption{\footnotesize  
Angular coverage of the multidetector array FAUST \cite{FAUST}.
}
\label{ffaust}
\end{figure}

Two-body interaction can be included in the framework of quantum kinetic 
theory as represented by the Boltzmann transport equation, which describes 
the reaction dynamics in terms of both mean-field and two-body dissipation. 
Various semi-classical methods featuring the concept of pseudo-particle  
are used to simulate the solution of the Boltzmann transport equation.  
The most common approximations are Vlasov-Uhling-Uhlenbeck ( VUU ) \cite{VUU}, 
Boltzmann-Uhling-Uhlenbeck ( BUU ) \cite{BUU}, Landau-Vlasov ( LV ) \cite{LV}, 
Boltz\-mann-Nordheim-Vlasov ( BNV ) \cite{BNV} and Quantum Molecular Dynamics 
( QMD ) \cite{QMD} methods. Results of such simulations typically offer a 
qualitative description of the main features of reaction dynamics including 
multifragmentation. Using the transport theory, the properties of the nuclear 
equation of state such as nuclear incompressibility and symmetry energy 
can be investigated phenomenologically. 

\begin{figure}[htbp]
\centering
\vspace{5mm}
\includegraphics[width=9.5cm,height=5.5cm]{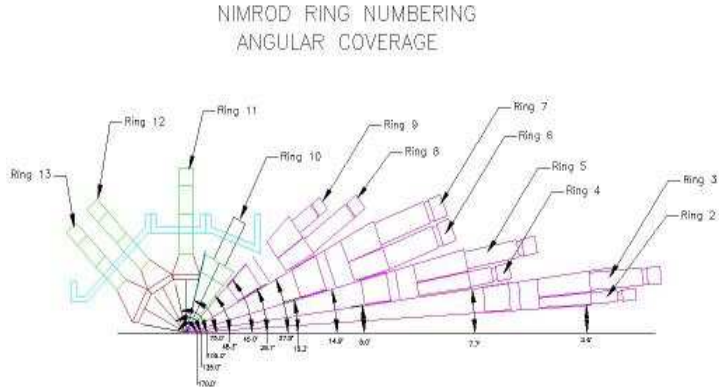}
\caption{\footnotesize  
Diagram of angular coverage of detector rings in the 4$\pi$ array NIMROD 
\cite{NIMROD}.
}
\label{fnimrod}
\end{figure}

\section*{Experimental aspects~ of~ multifragmentation studies}

Experimental studies of isotopic trends in nuclear multifragmentation were  
enabled by the recent progress in acceleration and detection techniques. 
A crucial condition is unambiguous isotopic identification of detected 
fragments, which can be achieved using high-resolution particle telescopes. 
In order to reconstruct the multifragmenting system and to understand the 
production mechanism of the hot nuclei it is desirable to detect the charged 
particles in a wide angular range, ideally in the 4$\pi$ geometry. In this 
section we briefly review the experimental devices ( using the facilities 
installed at Texas A\&M University as examples ) and essential methods 
relevant to the studies of isospin degrees of freedom in nuclear 
multifragmentation. 

\begin{figure}[t]
\centering
\vspace{5mm}
\includegraphics[width=9.0cm,height=7.0cm]{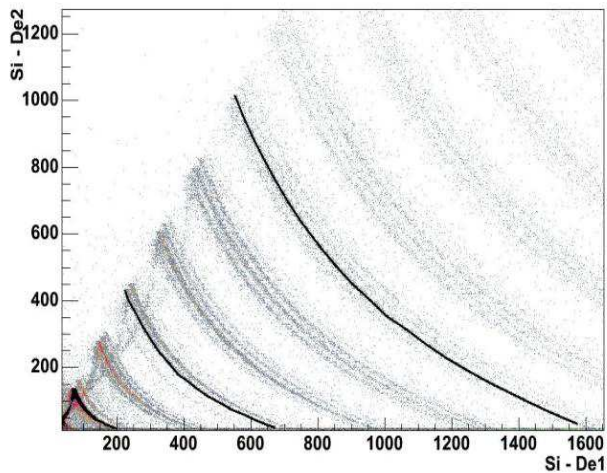}
\caption{\footnotesize  
An example of the raw Si-Si spectra  \cite{MVCal}. 
The $^{4}$He , $^{9}$Be and $^{15}$N lines used for calibration 
are shown as red lines. The data from reaction $^{124}$Sn+$^{124}$Sn
at 28 AMeV were collected in ring 8 of NIMROD. 
}
\label{fsisiraw}
\end{figure}
 
\subsection*{Arrays of charged particle detectors with large angular coverage}

During the process of multifragmentation, multiple charged particles 
are emitted in a wide angular range. Angular distribution of fragments 
in the laboratory frame is determined by the production mechanism 
of the hot nucleus and by the choice of the projectile and target nuclei. 
The design of arrays of charged particle detectors must be optimized in order 
to detect and resolve emitted charged particles from the expected 
multifragmentation source. This can be achieved by choice of angular coverage, 
granularity and detection thresholds.

A state-of-the-art forward array, the multidetector array FAUST \cite{FAUST}, 
installed at the Cyclotron Institute of 
Texas A\&M University, covers forward angles between 2 - 35 
$^{\circ}$ where the fragments originating from 
the projectile-like source can be expected. It consists of 68 charged particle 
telescopes arranged into 5 rings, as shown in Fig. \ref{ffaust}. 
Each particle telescope 
consists of 300 $\mu$m thick silicon detector followed by 3 cm thick CsI(Tl) 
crystal. Angular ranges covered by the rings are chosen in order to distribute 
the multiplicity of detected particles evenly, thus avoiding the cases when 
one telescope is hit by multiple particles. The mass and atomic number 
of the detected charged particles can be identified up to Z=6. Forward arrays 
such as FAUST are optimized for the study of projectile multifragmentation, 
specifically for asymmetric reactions where target nucleus is much heavier 
than the projectile.

In order to study symmetric reactions where a hot composite source is created 
it is necessary to cover largest possible angular range, ideally 4$\pi$. 
For instance, the multidetector NIMROD \cite{NIMROD} at the Cyclotron 
Institute of Texas A\&M 
University provides such a coverage using 96 detector modules arranged 
in 12 rings ( see Fig. \ref{fnimrod} ). Light charged particles ( Z$\leq$2 ) 
are detected and identified in the whole range while a subset of modules 
( 2 per ring ) offer isotopic identification up to Z=8. Furthermore, 
atomic number of heavy residues can be resolved using ionization chambers 
and neutron multiplicity can also be determined since the array is surrounded 
by a 4$\pi$ neutron multiplicity counter ( neutron ball.).  

\subsection*{Identification of isotopes}

\begin{figure}[t]
\centering
\vspace{5mm}
\includegraphics[width=9.0cm,height=7.0cm]{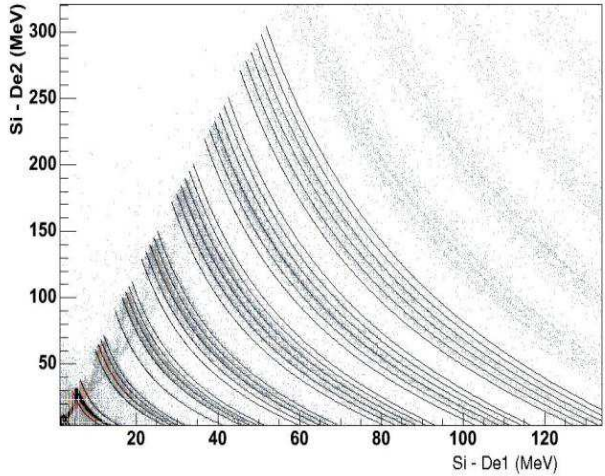}
\caption{\footnotesize  
Calibrated Si-Si spectrum from Fig. \ref{fsisiraw}. 
The isotopic lines used for identification are shown. 
}
\label{fsisi}
\end{figure}

An important step in the off-line analysis of data from the multidetector 
arrays with large angular coverage such as FAUST and NIMROD is to 
identify as many fragment species as possible. 
The method of isotope identification 
is based on well known particle telescope technique in which the isotopes 
are resolved in two-dimensional $\Delta$E-E spectra of energy losses 
in two detectors ( typically a thin one followed by a thick one ). 
We developed and successfully used \cite{MVCal} a method which enables 
to carry out 
identification and energy calibrations simultaneously using a minimization 
procedure applied to two-dimensional spectra. In the experimental spectra, 
the lines for three known isotopes ( typically the most characteristic 
isotopes such as $^{1}$H, $^{4}$He, $^{9}$Be ) are assigned and calibration 
is carried out by a minimization procedure where these lines are fitted 
to corresponding calculated energy losses for a given $\Delta$E-E telescope. 
The calibrations coefficients are thus obtained as optimum values 
of minimization parameters. 

\begin{figure}[h]
\centering
\vspace{5mm}
\includegraphics[width=9.0cm,height=7.cm]{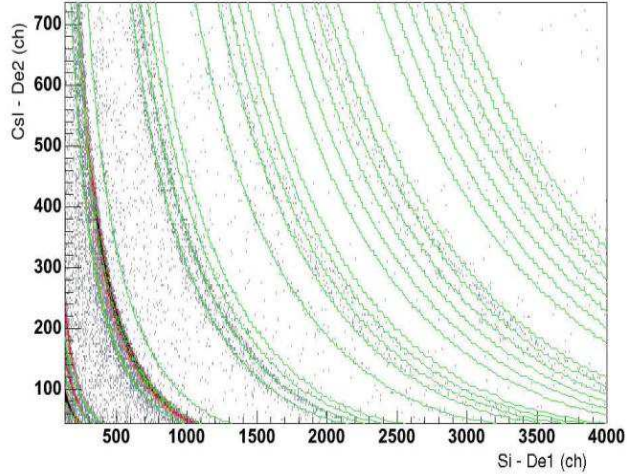}
\caption{\footnotesize  
An example of the raw Si-CsI spectra with de-calibrated isotopic 
dependences of energy losses. 
The $^{4}$He , $^{7}$Li and $^{9}$Be lines were used in calibration. 
The data from reaction $^{40}$Ca+$^{27}$Al
at 45 AMeV were collected in ring E of FAUST. 
}
\label{fsicsi}
\end{figure}

\begin{figure}[h]
\centering
\vspace{5mm}
\includegraphics[width=10.5cm,height=7.0cm]{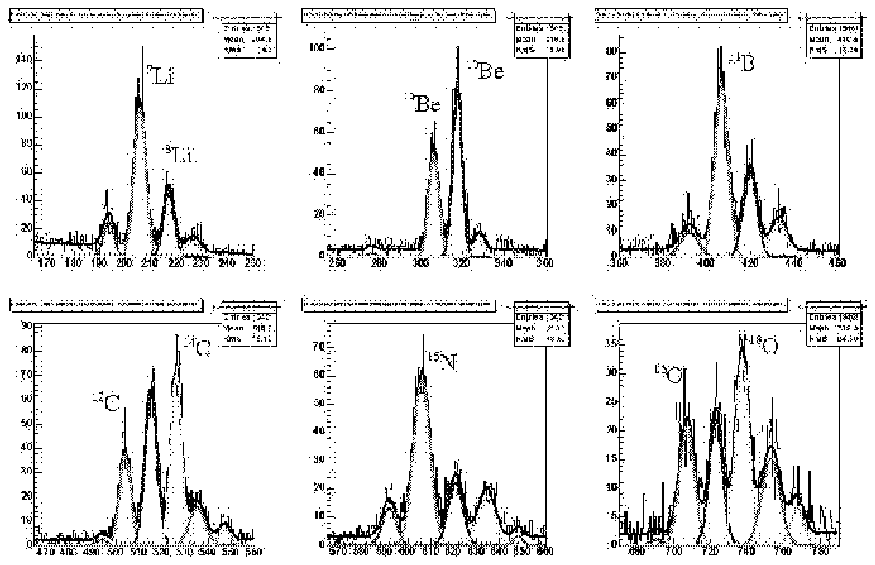}
\caption{\footnotesize  
Isotope resolution obtained after linearization of Si-Si spectra. 
Data from the reaction 
$^{124}$Sn+$^{124}$Sn at 28 AMeV were collected in the ring 8 of NIMROD 
\cite{DANF2001,Shetty}. }
\label{fgident}
\end{figure}

\begin{figure}[h]
\centering
\vspace{5mm}
\includegraphics[width=10.5cm,height=7.0cm]{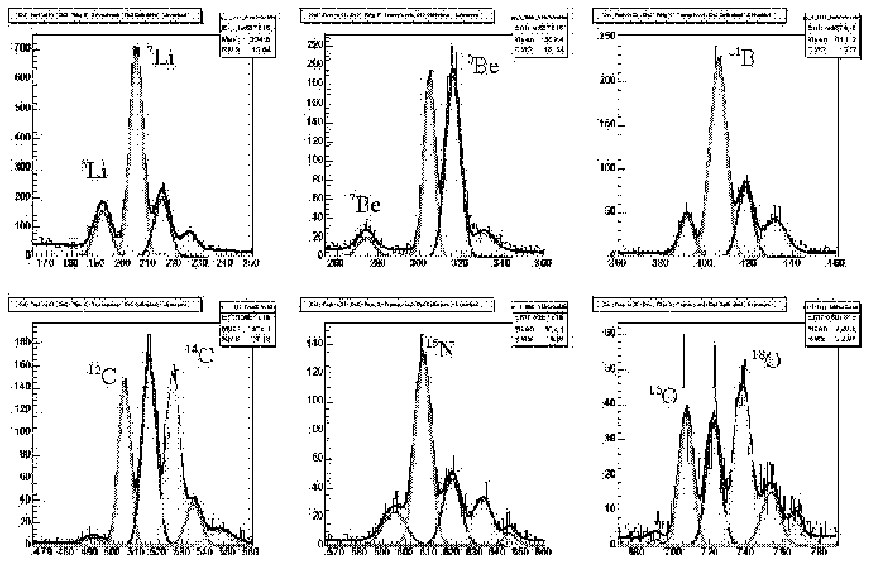}
\caption{\footnotesize  
Isotope resolution obtained after linearization of Si-CsI spectra. 
Data from the reaction 
$^{124}$Sn+$^{124}$Sn at 28 AMeV were collected in the ring 8 of NIMROD
\cite{DANF2001,Shetty}. }
\label{fgident2}
\end{figure}

For Si-Si calibration, the formula 
$E = a_0 + a_1 X + a_2 \sqrt{X}$ was used for both Si detectors 
( $X$ being the amplitude of electronic signal ). 
Additional constraints were applied in the minimization functional in order 
to position correctly the punch-through points. Figure \ref{fsisiraw} 
shows an example of the experimental spectra 
with selected $^{4}$He , $^{9}$Be and $^{15}$N lines  
and the calibrated spectrum with calculated $\Delta$E-E lines 
for a wide range of isotopes is shown in Fig. \ref{fsisi}. 
The data used  were collected in ring 8 of the NIMROD array in the reaction 
$^{124}$Sn+$^{124}$Sn at 28 AMeV. 
One can see that the overall agreement in the shape 
of calculated and calibrated energy losses  
is very good, both interpolating within and extrapolating beyond the range 
of selected isotopes ( at least to the neighboring element ). The way 
to identification of different isotopes by linearization of experimental 
spectra and comparison to calculated lines is straightforward. 

For Si-CsI(Tl) calibration a similar procedure can be  
carried out. The difficulty when compared to Si-Si telescope 
is caused by the fact that the measured signal in CsI(Tl) crystal 
is the light output which depends on particle energy non-linearly 
( including explicit dependence on particle mass and charge ).  
An approximate formula of Tassan-Got \cite{TGCal} for the energy 
calibration of CsI(Tl) detector was used

\begin{equation}
E = \sqrt { L^2 + 2 \rho L ( 1 + \ln (1+\frac{L}{\rho})) } 
\label{csical}
\end{equation}

\noindent 
where L is the light output and $\rho=\eta AZ^2$ where $\eta$ is 
a calibration parameter. An example of obtained agreement 
of experimental and calculated lines 
is given in Figure \ref{fsicsi} using the data detected in the FAUST array 
in the reaction  
$^{40}$Ca+$^{27}$Al at 45 AMeV. The calibration of Si detector was 
obtained using radioactive source of $\alpha$-particles 
and the calibration coefficients for CsI(Tl) detector were obtained 
using minimization procedure. Due to explicit mass and charge dependence 
the calibration must be preceded by identification, so in this case the  
calculated $\Delta$E-E lines are de-calibrated using calibration coefficients 
and projected onto experimental spectrum. 
Again the overall agreement is quite good for both interpolation and 
extrapolation.  

\begin{figure}[h]
\centering
\vspace{5mm}
\includegraphics[width=9.5cm,height=6.cm]{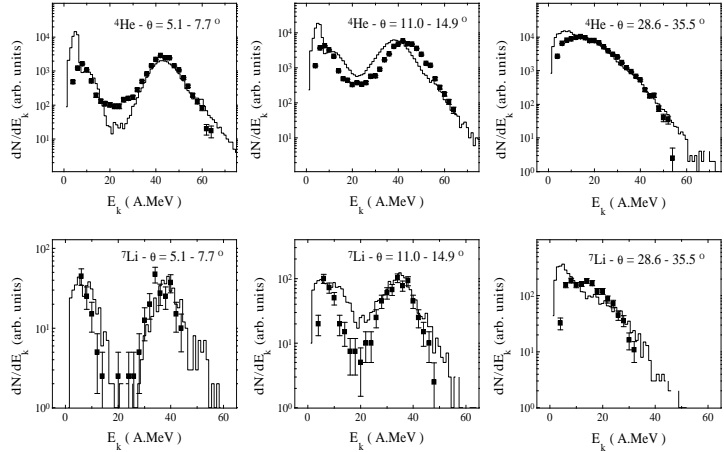}
\caption{\footnotesize  
The experimental particle energy spectra \cite{DANF2001} compared 
to the results of simulations \cite{Prod}. 
}
\label{fgspec}
\end{figure}

The examples of obtained isotope resolution ( after linearization ) 
are given in Figs. \ref{fgident},\ref{fgident2}. 
Data from the reaction 
$^{124}$Sn+$^{124}$Sn at 28 AMeV were collected in the ring 8 of NIMROD. 
One isotope line for each Z was used in linearization. 
The examples of particle spectra calibrated using the above methods are given  
in Fig. \ref{fgspec}. Data used in Fig. \ref{fgspec} were collected 
in the reaction $^{124}$Sn+$^{28}$Si at 28 AMeV using NIMROD 
\cite{DANF2001,Shetty}. 
The experimental spectra are compared to simulations \cite{Prod} 
described in previous section. 

\subsection*{Event characterization}

Modern multidetector arrays with large angular coverage offer a possibility 
to study not only the inclusive observables such as isotope yields and 
particle spectra but allow to study development of the fragment properties 
with evolving reaction dynamics. Typical goal of the exclusive 
study is to identify dynamical properties of the hot multifragmentation 
source being created in the early stage of the nucleus-nucleus collision. 
Here we provide some examples for both peripheral and violent collisions. 

\begin{figure}[htb]
\centering
\vspace{10mm}
\includegraphics[width=8.5cm,height=5.cm]{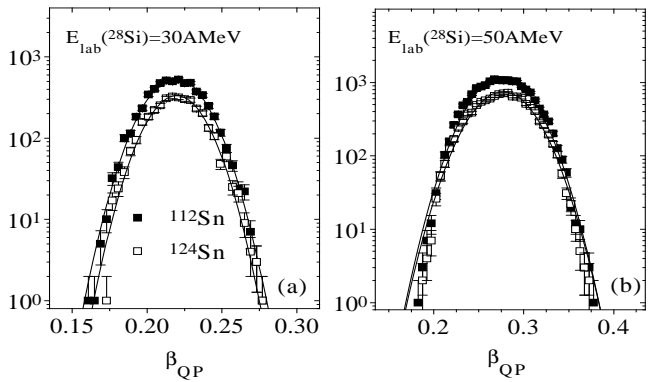}
\caption{ Experimental velocity distributions of the fully isotopically
resolved quasiprojectiles with $ Z_{tot}=12-15 $ \cite{SiSnNExch} 
( solid and open squares mean $^{112}$Sn and $^{124}$Sn target,
(a) and (b) the projectile energy
30 and 50 AMeV, respectively ).  Solid lines mean Gaussian fits.  }
\label{vqpex}
\end{figure}

Using the forward arrays like FAUST one 
typically observes the multifragmentation of the hot quasiprojectile 
which originates from peripheral and mid-peripheral nucleus-nucleus collision. 
We present here an example of the projectile multifragmentation 
in the reaction of $^{28}$Si beam with $^{112}$Sn and $^{124}$Sn targets 
\cite{SiSnNExch}. 
A hot multifragmenting source was identified using the calorimetry 
technique for the events  where 
all emitted fragments were isotopically identified. The 
velocity distributions of the reconstructed quasiprojectiles with total charge
$ Z_{tot}=12-15 $ were determined. 
Resulting velocity distributions for the projectile energies
30 and 50 AMeV are given in Fig. \ref{vqpex}a,b. 
Solid squares represent the reactions with $^{112}$Sn target
and open squares reactions with $^{124}$Sn. For a given
projectile energy, the mean velocities and widths of distributions
are practically identical for both targets. The shapes of velocity
distributions are close to Gaussian. 
Velocity distributions are symmetric and
no significant low or high energy tails are observed.
Thus, the reconstructed quasiprojectiles appear to originate from
a single source which can be identified with the quasiprojectile.

\begin{figure}[htb]
\centering
\includegraphics[width=7.5cm,height=7.cm]{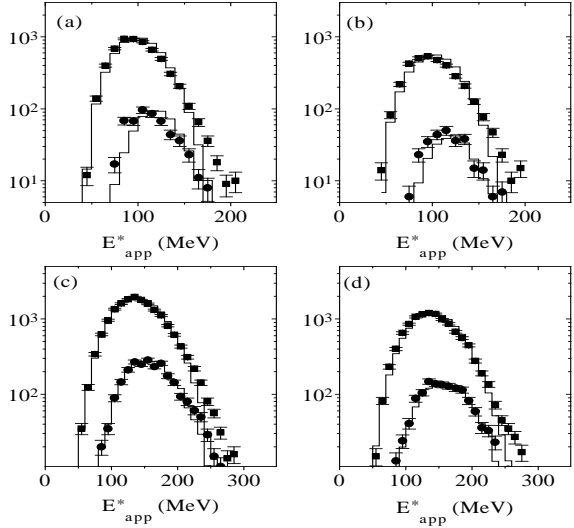}
\caption{ Distributions of the reconstructed apparent excitation energies
of the quasiprojectiles \cite{SiSnNExch}. Symbols mean experimental 
distributions of the set of fully isotopically resolved quasiprojectiles
with $ Z_{tot}=14 $ ( solid circles )
and $ Z_{tot}=12-15 $ ( solid squares ).
Solid histograms mean simulated distributions
for corresponding data sets, 
(a) - $^{28}$Si(30 AMeV)+$^{112}$Sn,
(b) - $^{28}$Si(30 AMeV)+$^{124}$Sn,
(c) - $^{28}$Si(50 AMeV)+$^{112}$Sn,
(d) - $^{28}$Si(50 AMeV)+$^{124}$Sn.}
\label{exsim}
\end{figure}

\begin{figure}[ht]
\centering
\includegraphics[width=9.5cm,height=6.cm]{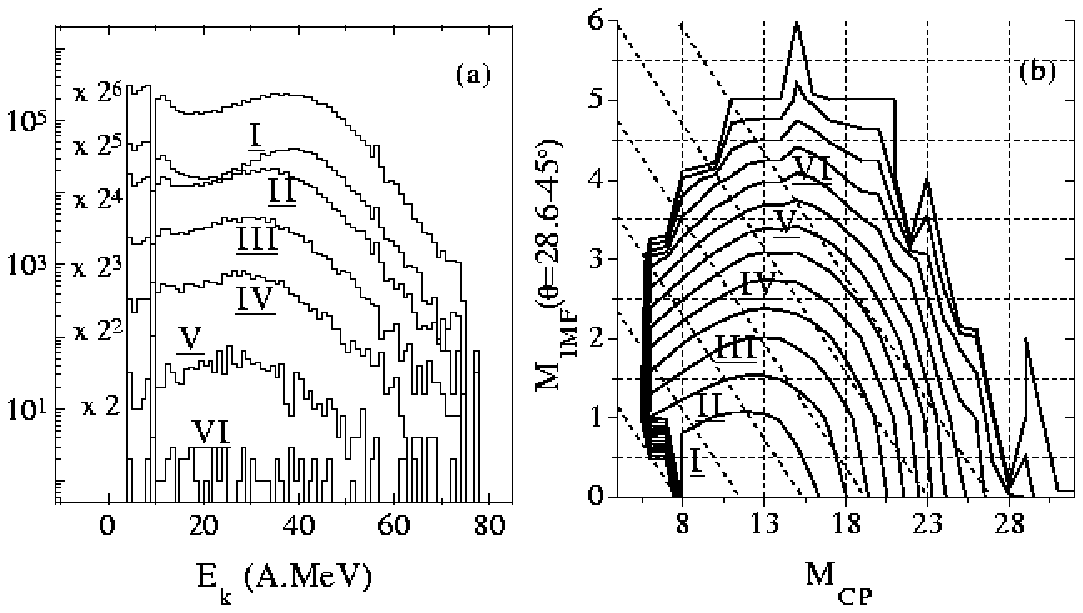}
\caption{ (a) - Inclusive spectrum of $\alpha$-particles ( top )
from reaction $^{124}$Xe+$^{124}$Sn and spectra for 6 centrality bins 
ordered from top to bottom by increasing centrality \protect\cite{DANF2001}.
(b) - Centrality cuts in the M$_{IMF}$ vs. M$_{CP}$ histogram.
Skewed lines indicate centrality cuts used in (a), horizontal
and vertical lines indicate one-dimensional cuts in
M$_{IMF}$ and M$_{CP}$ histograms.}
\label{nimspec}
\end{figure}

An apparent charged particle excitation energy
of the quasiprojectile can be reconstructed
for each projectile fragmentation event from the energy balance
in the center of mass frame of the quasiprojectile

\begin{equation}
 E_{app}^{*}=\sum _{f}(T_{f}^{QP}+\Delta m_{f})-\Delta m_{QP}
\label{exqp}
\end{equation}

where $ T_{f}^{QP} $ is the kinetic energy of the fragment in the
quasiprojectile frame and $ \Delta m_{f} $ 
and $ \Delta m_{QP} $ are mass 
excesses of the fragment and quasiprojectile, respectively. Even if emitted
neutrons are not included in this observable, it is useful for the light 
fragmenting systems, where the neutron emission is not expected 
to dominate. The distributions of
the apparent quasiprojectile excitation energies
reconstructed from fully isotopically resolved events are shown 
in Fig. \ref{exsim} for both $ Z_{tot}=14 $ and $ Z_{tot}=12-15 $  
( solid circles and squares respectively ) along with the 
distributions obtained using DIT/SMM simulation \cite{SiSnNExch} 
( solid histograms ). 
The agreement of the simulated and experimental apparent quasiprojectile
excitation energy distributions is quite good. The onset of multifragmentation 
into channels with
$ Z_{f} \leq 5 $ in the low energy part is described with good precision
for all sets of data.

While it is possible to fully reconstruct the projectile-like sources 
in the mass range A=20-30 in peripheral collisions, 
with increasing violence of the nucleus-nucleus collision 
a hot nucleus is formed from the parts of both the projectile and target. 
The excitation energy of such a source is high enough for emission of several 
intermediate mass fragments ( IMF ). Information on both mass and charge of
emitted IMFs can be crucial to determine the emission 
mechanism and thus properties of the emitting source. 
Various techniques have been developed 
in order to characterize the impact parameter of the collision 
( for a detailed study see \cite{Frank} ). A simple method which 
is nevertheless widely used to determine the centrality of collision 
is based on the assumption that the particle multiplicity increases 
monotonously with impact parameter \cite{Cavata}. However, such assumption 
may be too simplistic and more sophisticated methods using the combinations 
of several observables are developed \cite{Frank}. As an example, 
we present here a method \cite{DANF2001} where the centrality bins  
are selected in the two-dimensional histogram of IMF multiplicities 
( M$_{IMF}$ ) vs multiplicities of charged particles ( M$_{CP}$ )
by parallel lines chosen so that the most central bin integrates   
the events with highest multiplicities of charged particles and 
highest IMF multiplicities ( see Fig. \ref{nimspec}b ) 
with the intermediate event classes, thus reflecting the effect of finite size. 
Fig. \ref{nimspec}a shows that the shapes of spectra develop 
with increasing centrality.

\begin{figure}[ht]
\centering
\includegraphics[width=7.5cm,height=7.cm]{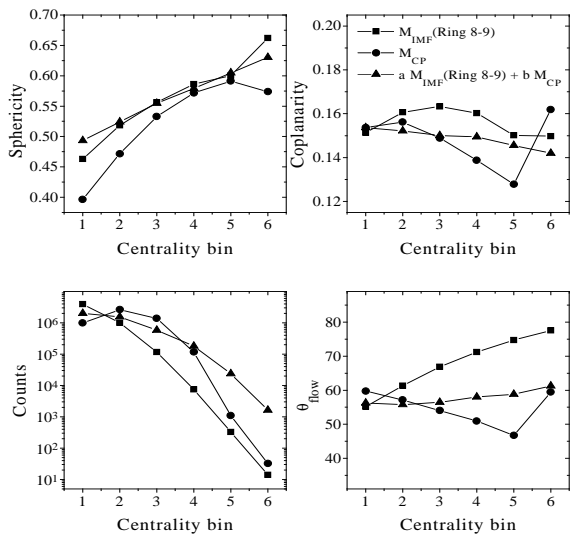}
\caption{ 
Test of centrality selections in terms of sphericity (a), coplanarity (b), 
event statistics (c) and flow angle (d) for two-dimensional M$_{IMF}$ 
vs. M$_{CP}$ cuts ( triangles ) as well as for the one-dimensional 
M$_{IMF}$ ( squares ) and M$_{CP}$ ( circles ) cuts \cite{DANF2001}. 
}
\label{fgcnt}
\end{figure}

Event shape analysis was carried out for two-dimensional M$_{IMF}$ 
vs. M$_{CP}$ cuts as well as for the one-dimensional M$_{IMF}$ and 
M$_{CP}$ cuts. All types of centrality cuts exhibit similar trends
in terms of sphericity and coplanarity ( see Fig. \ref{fgcnt} ), 
nevertheless some differences can be observed in the most central bins. 
When considering the total number of events selected by 
various criteria, the statistics in the most central bin of the 
two-dimensional selection increased by two orders
of magnitude compared to the traditional M$_{CP}$ selection 
( see Fig. \ref{fgcnt} ). 
Such a gain in statistics is caused by better separation of central and 
peripheral events. The one-dimensional cuts typically admix high-multiplicity 
peripheral events to a smaller number of genuine central events, thus 
decreasing the overall centrality. The use of two-dimensional  
criteria allows more detailed studies of central events. 

\section*{Isotopic trends}

The isotopic composition of the multifragmentation products such as light
charged particles ( LCP ), intermediate mass fragments (IMF ) and 
heavy residues is an 
experimental information which can be used to trace the dynamics of isospin 
degrees of freedom during the multifragmentation and, using grand-canonical 
picture, to extract the values of thermodynamical observables of the system
undergoing multifragmentation such as temperature and free nucleon densities. 
The isotopic composition can be studied both as a bulk or in terms of 
yield ratios of specific isotopes. A variation of the neutron content 
of reaction products can be used to characterize the response 
of the isotopic composition to variation of initial conditions. 

\subsection*{Isotopic distributions} 

\begin{figure}[ht]
\centering
\includegraphics[width=9.cm,height=10.cm]{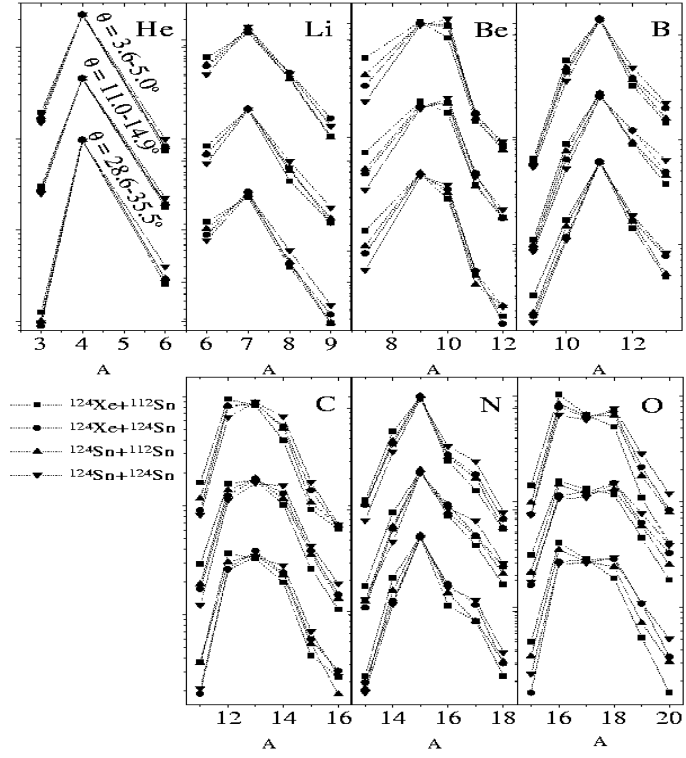}
\caption{ 
Isotopic distributions of fragments with Z=2-8 from 
four reactions of $^{124}$Sn,$^{124}$Xe beams with $^{112,124}$Sn 
targets at 28 \hbox{AMeV} \cite{DANF2001,Shetty}, measured using 
the 4$\pi$ multi-detector array NIMROD \cite{NIMROD} at three angular ranges.  
}
\label{fgyazn}
\end{figure}

A wide variety of fragment species is typically produced in multifragmentation 
of heavy nuclei. In order to obtain information on isospin dynamics, 
an obvious choice is to study the behavior of isotopic chains. 
As an example we provide in Fig. \ref{fgyazn} isotopic distributions 
of fragments with Z=2-8 from 
four reactions of $^{124}$Sn,$^{124}$Xe beams with $^{112,124}$Sn 
targets at 28 \hbox{A MeV} \cite{DANF2001,Shetty} studied at
the Cyclotron Institute of Texas A\&M University
using the 4$\pi$ multi-detector array NIMROD \cite{NIMROD}. 
The isotopic distributions from different reactions are distinguished by 
different symbols. Each panel shows the isotopic distributions measured 
at three angular ranges corresponding to rings 2, 5 and 8 ( from top 
to bottom ). 

As one can see, the overall shapes remain quite similar at different 
angular ranges. It is interesting to note that the isotopes with maximum 
yield for Z$\geq$3 are typically those with A=2Z+1 for both even and odd 
Z. The effect of different reactions is most pronounced 
at the proton and neutron-rich tails of distributions, neutron-rich 
isotopes being more abundant in neutron-rich systems and vice versa, 
while the central part corresponding to $\beta$-stable nuclei is relatively 
insensitive. In any case, such comparisons of the shapes of isotopic 
distributions are rather qualitative. A quantitative analysis can be 
carried out in terms of integral observables such as momenta of distributions.

\begin{figure}[ht]
\centering
\includegraphics[width=9.cm,height=6.cm]{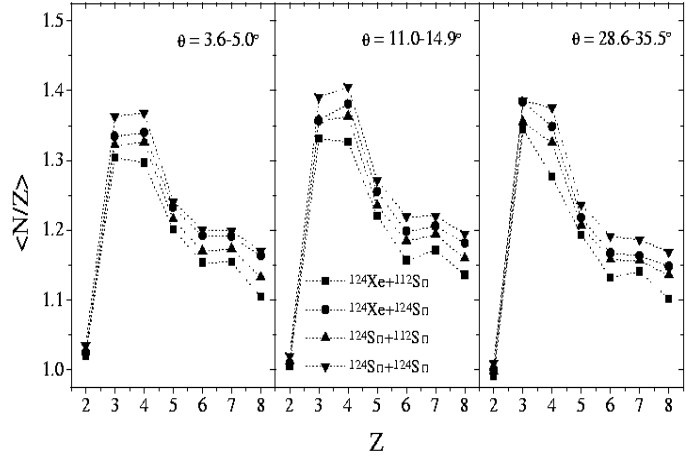}
\caption{ Mean N/Z-ratios of emitted LCPs and IMFs with Z=2-8 from
four reactions obtained at different angles \protect\cite{DANF2001,Shetty}.}
\label{fgyz}
\end{figure}

To express the effect of experimental conditions such as 
projectile-target combination or detection angle 
quantitatively one can evaluate the mean N/Z- 
( neutron to proton ) ratio for each isotopic chain. 
In Fig. \ref{fgyz} are given the mean N/Z-ratios of emitted LCPs and IMFs
with Z=2-8 from four reactions obtained at different angles \cite{DANF2001}.
The overall dependence is similar at all angular ranges.
He-isotopes are dominated by $\alpha$-particles.
Mean N/Z ratios are highest for Li- and Be-isotopes 
and decrease gradually with increasing atomic number of IMFs.
When comparing mean N/Z ratios from four reactions 
two characteristic patterns can be distinguished. At most forward
angles the mean N/Z ratios of Li- and Be-isotopes appear
to track with the isospin asymmetry of the composite projectile-target
system what leads to grouping of experimental point into a typical 1-2-1 
pattern. On the other hand
the isotopes with Z=5-8 appear to track with the isospin asymmetry
of the target nucleus what creates a 2-2 pattern. At angles 11.0-14.9$^{\circ}$
the 1-2-1 pattern can be identified for Li- and Be-isotopes
while for heavier fragments the patterns vary.
At angles 28.6-35.5$^{\circ}$ one can recognize the 2-2 pattern
for Li-isotopes while at heavier fragments the 1-2-1 pattern dominates.
In principle, one can identify the 1-2-1 pattern with the fragments
emitted from a hot composite source and 2-2 pattern to the emission from
a quasiprojectile. The fragments with Z=5-8 at forward angles 
possibly originate from the binary de-excitation channels of the 
quasiprojectile while the 2-2 pattern of Li-isotopes at the most
central angles suggests either backward emission from the 
quasiprojectile or emission from quasitarget. Similar isospin asymmetry
patterns can be identified also in the isotopic yield
distributions in Fig. \ref{fgyazn}. Thus, the mean values of N/Z ratio 
appear to exhibit an interesting information on isospin dynamics. 
The relative variation of the mean N/Z values is nevertheless rather small 
when compared to the variation of isospin-asymmetry of the projectile-target 
combination and more sensitive isospin tracers should be identified. 

\subsection*{Yield ratios}

As demonstrated in the previous subsection, variation of the projectile 
and target nuclei influences the integral 
characteristics of isotopic distributions such as mean values 
of N/Z ratio rather weakly. Nevertheless, when observing 
the evolution of isotopic distributions in different reactions 
and angular ranges, one can notice that the relative yields 
of neighboring isotopes are rather sensitive ( especially at the 
tails of distributions ). Such yield ratios have been widely used 
in experimental studies of multifragmentation data during last 15 years 
\cite{YRat}. 

\begin{figure}[ht]
\centering
\includegraphics[width=9.cm,height=6.cm]{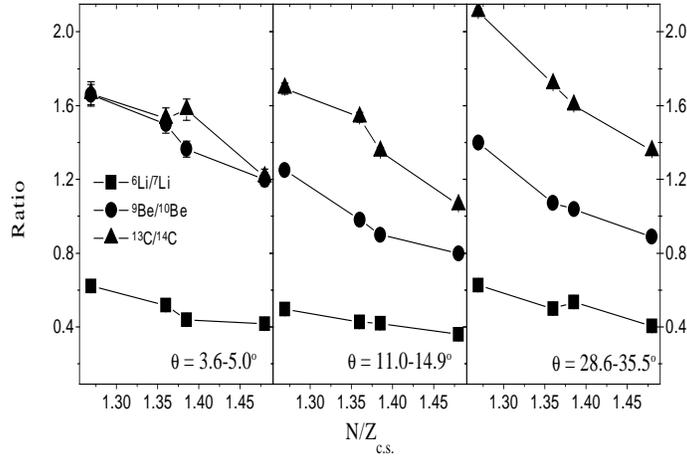}
\caption{ Three 
isotopic yield ratios for the multifragmentation data from 
four reactions of $^{124}$Sn,$^{124}$Xe beams with $^{112,124}$Sn 
targets at 28 \hbox{AMeV} \cite{DANF2001,Shetty} 
at three different angular ranges. 
The data from different reactions are characterized by a N/Z-ratio 
of the composite system \hbox{( N/Z$_{c.s.}$ ).} }
\label{fgisotop}
\end{figure}

As an example, we present in Fig. \ref{fgisotop} the values of three isotopic 
yield ratios, again for the multifragmentation data \cite{DANF2001,Shetty} from 
four reactions of $^{124}$Sn,$^{124}$Xe beams with $^{112,124}$Sn 
at 28 \hbox{AMeV}. In most cases one can observe 
an approximately exponential dependence of the isotopic yield ratio 
on the N/Z ratio of the composite system ( N/Z$_{c.s.}$ ), which can be 
expected according to formula (\ref{eqnynz}) when assuming linear dependence 
of neutron chemical potential on N/Z$_{c.s.}$. The yield ratios 
depend rather strongly on the angular range. 
Such a dependence can be caused by different isospin dynamics 
at different angles. 

\begin{figure}[ht]
\centering
\includegraphics[width=9.cm,height=6.cm]{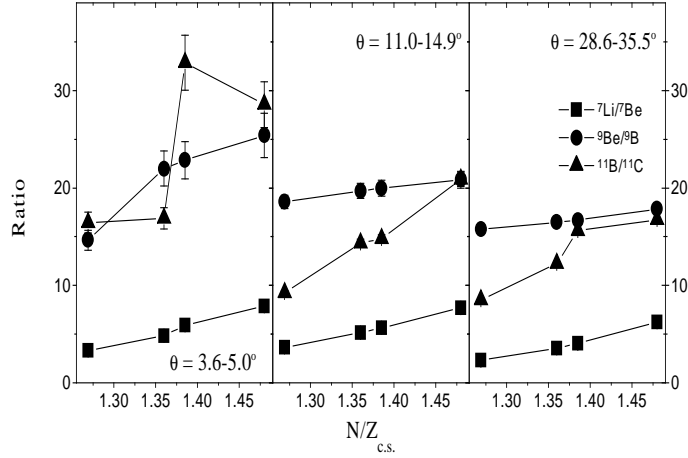}
\caption{ Three 
isobaric yield ratios for the multifragmentation data from 
four reactions of $^{124}$Sn,$^{124}$Xe beams with $^{112,124}$Sn 
targets at 28 \hbox{AMeV} \cite{DANF2001,Shetty} 
at three different angular ranges. 
The data from different reactions are characterized by a N/Z-ratio 
of the composite system ( N/Z$_{c.s.}$ ). }
\label{fgisobar}
\end{figure}

Furthermore, we present in Fig. \ref{fgisobar} the values of three 
isobaric yield ratios for the same multifragmentation data. Specifically, 
the isobaric ratios are evaluated for pairs of mirror nuclei. The 
sensitivity of isobaric ratios to N/Z$_{c.s.}$ is significantly stronger 
than in the previous case, especially for the ratios $^{9}Be/^{9}B$ 
and $^{11}B/^{11}C$  where the fragments $^{9}B$ and $^{11}C$ are populated 
rather weakly. The overall behavior is nevertheless rather irregular. 
Especially 
the observed pattern at most forward angles suggests a contribution from 
the cold process such as projectile break-up, which appears to 
contribute considerably to production of isospin-asymmetric species 
at forward angles. Such observation is in agreement with available 
experimental data \cite{ColdBkup}. 

\begin{figure}[ht]
\centering
\includegraphics[width=9.cm,height=9.cm]{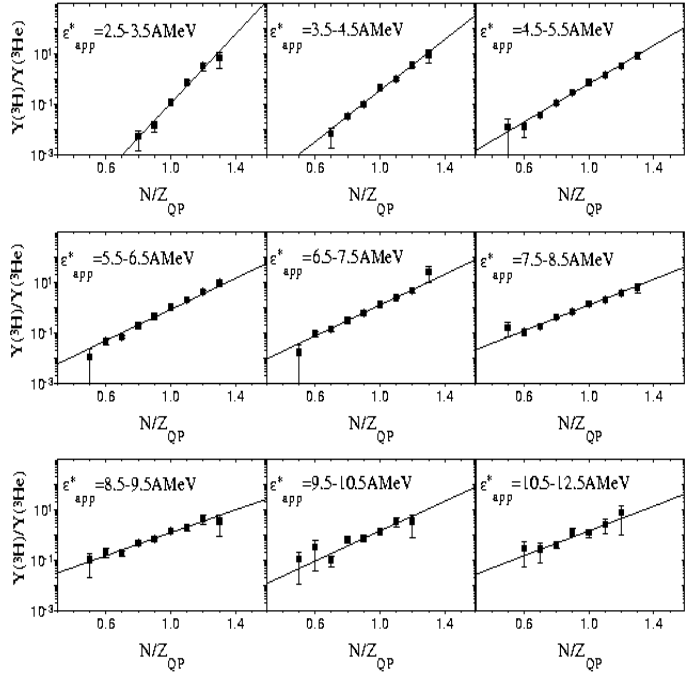}
\caption{
Dependence of the yield ratio Y($^{3}$H)/Y($^{3}$He) on N/Z ratio of the
isotopically resolved quasiprojectiles with $Z_{QP}=12-15$ for
nine bins of $\epsilon^{*}_{app}$ \protect\cite{MuTemp}. }
\label{fgy3rat}
\end{figure}

A behavior of the yield ratio of mirror nuclei $^{3}H/^{3}He$ was 
investigated in the quasiprojectile multifragmentation of a $^{28}$Si beam with 
$^{112}$Sn and $^{124}$Sn targets at 30 and 50 \hbox{AMeV}  \cite{MuTemp}.  
As described in experimental section, the properties of the system which 
actually undergoes multifragmentation, including N/Z ratio ( $N/Z_{QP}$ ), 
are known with good precision. 
Fig. \ref{fgy3rat} shows dependence of the isobaric yield 
ratio Y($^{3}$H)/Y($^{3}$He) on $N/Z_{QP}$ for nine bins of the apparent
excitation energy per mass unit of the quasiprojectile
( $\epsilon^{*}_{app}$ ). 
The experimental data are represented as squares
and the lines represent the exponential fits. 
The observed exponential dependence in all 
excitation energy bins is in agreement with formula (\ref{eqnynz}).  
The slopes are steepest at low
excitation energies and become flatter with increasing excitation energy. 

The apparent compliance of the yield ratio behavior to the grand-canonical 
picture as represented by the formula (\ref{eqnynz}) 
suggests the possibility to extract the values of thermodynamic observables,  
in particular temperature. The procedure for extraction 
of the temperature was suggested by Albergo \cite{Albergo}. When expressing the 
isotopic yield within the equilibrium limit of the grand-canonical ensemble, 
a simplified analogue of the formula (\ref{eqnynz}) is obtained \cite{KoonRand}

\begin{equation}
        Y(N,Z) = F(V,T) g(N,Z) 
	\exp((1/T)(B(N,Z)+N \mu_{n} + Z \mu_{p}))
\label{ynzalb}
\end{equation}

The isotopic yield is thus essentially governed by free neutron and proton 
chemical potentials $\mu_{n,p}$ , temperature $T$ and fragment binding energy 
$B(N,Z)$. The factor $F(V,T)$ encompasses information on the global properties 
of the fragment partition. The ground state spin degeneracy factor $g(N,Z)$ 
represents the internal state sum of the fragment, which is restricted to 
ground state only. In oder to extract temperature, a double isotope ratio 
is constructed \cite{Albergo} 

\begin{eqnarray}
        \frac{Y(N_1+1,Z_1)/Y(N_1,Z_1)}{Y(N_2+1,Z_2)/Y(N_2,Z_2)} 
=  \frac{g(N_1+1,Z_1)/g(N_1,Z_1)}{g(N_2+1,Z_2)/g(N_2,Z_2)} \times \nonumber \\
	\times \exp(\frac{B(N_1+1,Z_1)-B(N_1,Z_1)-B(N_2+1,Z_2)+B(N_2,Z_2)}{T})
\label{diralb}
\end{eqnarray}

and the temperature $T$ can be obtained as 

\begin{equation}
        T = \frac {\Delta B_{1234}} 
	{\ln{a \frac{Y(N_1+1,Z_1)/Y(N_1,Z_1)}{Y(N_2+1,Z_2)/Y(N_2,Z_2)}}} 
\label{tempalb}
\end{equation}

where 	$a=\frac{g(N_2+1,Z_2)/g(N_2,Z_2)}{g(N_1+1,Z_1)/g(N_1,Z_1)} $ and 
$\Delta B_{1234} = B(N_1+1,Z_1)-B(N_1,Z_1)-B(N_2+1,Z_2)+B(N_2,Z_2)$. Thus an 
estimate of the grand-canonical temperature can be determined using 
the isotope yields and some of their ground state characteristics. 

\begin{figure}[ht]
\centering
\includegraphics[width=12.cm,height=6.cm]{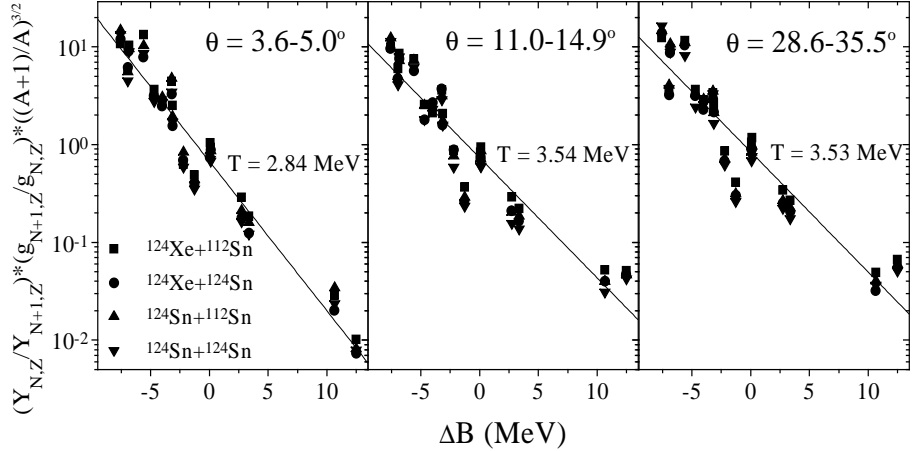}
\caption{ Systematics of corrected isotopic ratios $Y_{N,Z}/Y_{N+1,Z}$ from
four reactions plotted against the difference of binding energies
 \protect\cite{DANF2001}. }
\label{nimtemp}
\end{figure}

The formula (\ref{tempalb}) was used extensively in experimental studies 
during the last decade \cite{SpecMFrg,AlbExp}. Typically, the temperature 
determined using 
various thermometers ( sets of isotope yields ) exhibited large spread, 
possibly due to side-feeding from secondary decay. A method of correction 
for secondary decay was proposed by Tsang et al. \cite{AlbCorr}. 
In \cite{DANF2001} we employed 
a graphical method to extract the average temperature. The isotopic yield 
ratios, corrected for ground state spin degeneracy, are plotted as a function 
of binding energy difference ( see Fig. \ref{nimtemp} ). The data from four 
reactions of $^{124}$Sn,$^{124}$Xe beams with $^{112,124}$Sn targets 
at 28 \hbox{A MeV}, 
used already in Figs. \ref{fgisotop},\ref{fgisobar}, 
are investigated.  Individual double isotope ratio thermometers can be 
represented ( in the logarithmic scale ) when connecting 
the two points corresponding to used isotope ratios by a straight line. 
The inverse of the slope of such line is then the temperature. 
When assuming that the effect of side-feeding in the secondary decay 
is distributed randomly, an average temperature can be obtained 
by a linear fit of the systematics of all isotopic yield ratios. 

An information on the temperature can be obtained also from the 
slope of dependence of the isobaric yield ratio Y($^{3}$H)/Y($^{3}$He) 
on neutron to proton ratio of the quasiprojectile $N/Z_{QP}$.  
The data from reactions of a $^{28}$Si beam 
with $^{112}$Sn and $^{124}$Sn targets at 30 and 50 \hbox{A MeV} 
( see Fig. \ref{fgy3rat} ) were used \cite{MuTemp} and 
the extracted temperatures 
compared well with the temperatures obtained using the double 
isotope ratio thermometer ($^{2}$H/$^{3}$H)/($^{3}$He/$^{4}$He). 
An assumption was employed that the values of chemical potentials 
can be approximated by the values of separation energies. The reason 
why such approximation is applicable for hot expanded nucleus 
can be in the possible interplay of the effect of expansion 
( reducing absolute values of chemical potentials ) and 
separation of the isospin-asymmetric free nucleon gas from the 
isospin-symmetric heavier liquid phase \cite{MuTemp}, as indicated by observed 
inhomogeneous isospin distribution among light charged particles 
and heavier fragments \cite{IsoDist}. 

The isotopic yield can be, within the equilibrium limit of the grand-canonical 
ensemble, expressed in terms of the density of the free nucleon gas 
\cite{Albergo,KoonRand}

\begin{equation}
        Y(N,Z) \propto V \rho_n^N \rho_p^Z g(N,Z) 
	\exp(B(N,Z)/T)
\label{ynzrho}
\end{equation}

where $\rho_{n,p} \propto \exp(\mu_{n,p}/T)$. The values of yield ratios 
of mirror nuclei with $N-Z=1$ then can be expressed as   

\begin{equation}
        \frac{Y(N,Z)}{Y(Z,N)} \simeq \frac{g(N,Z)}{g(Z,N)} 
	\frac{\rho_n}{\rho_p} \exp(\Delta B/T) 
\label{yrmirr}
\end{equation}

thus relating the yield ratio to the density ratio of the free neutron and 
proton gas. The yield ratios of the mirror nuclei $^{3}H/^{3}He$, 
$^{7}Li/^{7}Be$ and $^{11}B/^{11}C$ were used in work \cite{Xu} 
to determine the ratio $\rho_n/\rho_p$ by an exponential extrapolation 
toward $\Delta B = 0$, yielding the values approximately 2 - 5. 
While such approximation is valid for grand-canonical ensemble 
of infinite systems, when dealing with the finite systems such as 
nuclei the values of free nucleon chemical potentials depend on 
the fragment partition in a way guaranteeing conservation 
of the mass and charge of the system. Thus the value of the 
ratio $\rho_n/\rho_p$ may differ strongly for different event 
classes and no globally applicable value may exist in the finite system. 
Furthermore, the effect of secondary emission may also contribute 
to uncertainty in determination of $\rho_n/\rho_p$. 

\subsection*{Isoscaling}

As shown in the previous subsection, the fragment yield ratios can be 
used to extract thermodynamical observables of the fragmenting system 
such as temperature and chemical potential. In the context of isotopic 
distributions, the fragment yield ratios represent the details of the 
distribution sensitive to isospin degrees of freedom. Similar sensitivity 
can be explored globally by investigating the ratio of isotopic yields 
from two processes with different isospin asymmetry, essentially dividing the 
two isotopic distributions in point-by-point fashion. When employing 
the formula (\ref{ynzalb}), such a ratio will depend on N and Z 
as follows \cite{TsangIso}

\begin{equation}
     R_{21}(N,Z) = Y_{2}(N,Z)/Y_{1}(N,Z) = C \exp(\alpha N  + \beta Z)
\label{r21isots}
\end {equation}       

where  $\alpha$ = $\Delta \mu_{n}$/T 
and $\beta$ = $\Delta \mu_{p}$/T, $\Delta \mu_{n}$ and $\Delta \mu_{p}$ 
are the differences in the free neutron and proton chemical potentials 
of the fragmenting systems. C is an  overall normalization constant.
Alternatively \cite{BotvIso} the dependence can be expressed as

\begin{equation}
     R_{21}(N,Z) = Y_{2}(N,Z)/Y_{1}(N,Z) 
     = C \exp(\alpha^{\prime} A  + \beta^{\prime} (N-Z) )
\label{r21isobt}
\end {equation}       

thus introducing the parameters which can be related to the isoscalar 
and isovector components of free nucleon chemical potential since 
$\alpha^{\prime}$ = $\Delta (\mu_{n}+\mu_{p})$/2T 
and $\beta^{\prime}$ = $\Delta (\mu_{n}-\mu_{p})$/2T. 

\begin{figure}[ht]
\centering
\includegraphics[width=12.cm,height=6.cm]{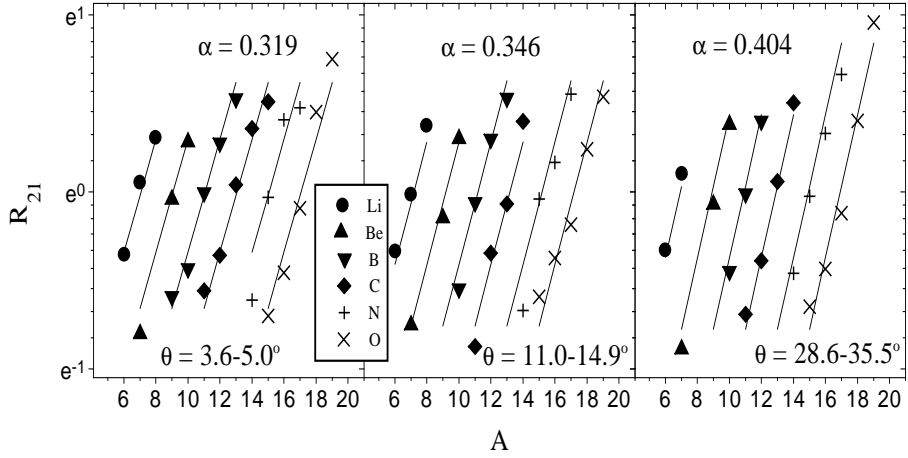}
\caption{ The isoscaling plots for the multifragmentation data from 
reactions of $^{124}$Sn+$^{124}$Sn and $^{124}$Xe+$^{112}$Sn 
at 28 \hbox{A MeV} \cite{DANF2001,Shetty} 
at three different angular ranges. }
\label{fgnimiso}
\end{figure}

An exponential scaling of $R_{21}$ with the isotope neutron and proton numbers 
was observed experimentally in the multifragmentation data 
from the  reactions of high-energy light particle with massive target nucleus 
\cite{Lozhkin,BotvIso} or from the reactions of mass symmetric projectile 
and target at intermediate energies \cite{TsangIso} and such 
behavior is called isoscaling \cite{TsangIso} ( the parameters 
$\alpha, \beta, \alpha^{\prime}, \beta^{\prime}$ being called isoscaling 
parameters ). Isoscaling behavior was further reported 
in the heavy residue data \cite{GSHRIso} and 
also in fission data \cite{Fisiso}. It was shown that the 
values of isoscaling parameters can be related to symmetry energy 
\cite{TsangIso,BotvIso}, to the level of isospin equilibration \cite{GSHRIso} 
and to the values of transport coefficients \cite{Fisiso}. An example 
of isoscaling behavior of the fragments with Z=3-8 detected in the 
reactions of $^{124}$Sn+$^{124}$Sn and $^{124}$Xe+$^{112}$Sn 
at 28 \hbox{AMeV} \cite{DANF2001,Shetty} is given in Fig. \ref{fgnimiso}.  
The isoscaling behavior is rather regular at all three angular ranges, 
the isoscaling parameter $\alpha$ increases with the detection angle. 
Such a trend of the isoscaling parameter $\alpha$ can be possibly related 
to the fact that with increasing angle the fragments originate 
from still more damped collisions, with increasing level of isospin 
equilibration between projectile- and target-like fragment \cite{GSNZEq}. 

\begin{figure}[h]                                        

\centering
\includegraphics[width=8.0cm, height=8.0cm ]{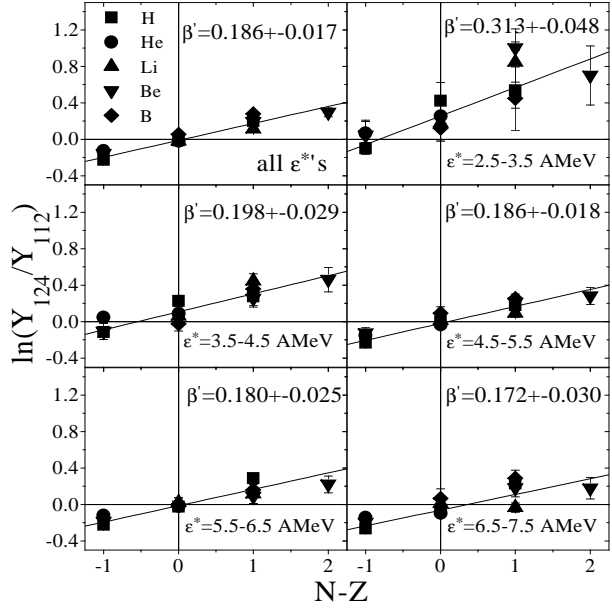}

\caption{
The isoscaling plots from the reactions of $^{28}$Si+$^{124,112}$Sn 
at 50 AMeV for the full set of isotopically resolved quasi-projectiles 
and for five bins of quasi-projectile excitation energy per nucleon
\cite{SiSnIso}. 
           }
\label{Iso50}
\end{figure}

In Fig. \ref{Iso50} are presented the isoscaling data 
from statistical decay of hot quasiprojectiles from the reactions 
$^{28}$Si+$^{124,112}$Sn at projectile energy 50 AMeV \cite{SiSnIso}. 
Observed charged 
particles with Z$\leq$5 were isotopically resolved and 
total observed charge was close to the charge of the projectile ( Z=12-15 )
\cite{RLSiSn,SiSnNExch}. Due to the restricted set of identified 
fragment species, the isoscaling plots are presented, according 
to formula (\ref{r21isobt}), as a function of isospin asymmetry. The isoscaling 
plots are presented not only for the full data but also for five 
bins of excitation energy. The slopes depend on the excitation energy 
in a similar way 
as the slope of the dependence of isobaric ratio Y($^{3}$H)/Y($^{3}$He) 
on the quasi-projectile N/Z observed in \cite{MuTemp} ( see 
Fig. \ref{fgy3rat} ). When compared to similar results at 30 AMeV, 
the isoscaling parameters $\beta^{\prime}$ for individual excitation energy 
bins do not depend on the projectile energy and the difference of the slopes 
of the inclusive data is caused by different excitation 
energy distributions of the hot quasi-projectiles at two projectile energies 
( see Fig. \ref{exsim} ). 

As demonstrated, isoscaling appears to be a global feature of the 
multifragmentation data and the values of isoscaling parameters show 
sensitivity to properties of the hot source. 
It is of interest to explore such a sensitivity in the studies 
of thermodynamical properties of the multifragmenting source.

\section*{Thermodynamical properties of the hot quasi-projectiles}

In the previous section we presented an overview of the methods by which 
the isotopic trends of the fragment data can be used 
to extract thermodynamical observables 
of the hot multifragmenting source. Here 
we present a systematic investigation of thermodynamical 
properties of the well characterized hot quasi-projectiles with 
mass A=20-30, originating from the reactions 
$^{28}$Si+$^{124,112}$Sn at projectile energy at 30 and 50 AMeV 
\cite{SiSnNExch}. 
Signatures and argumentation supporting the conclusion that multifragmentation 
of such quasiprojectiles is a statistical process will be presented. 
Furthermore, the signals for the isospin-asymmetric liquid-gas phase 
transition in the nuclear matter as a process underlying to multifragmentation 
will be investigated and implications for the nuclear equation of state 
at sub-saturation densities will be discussed. 

\subsection*{Production mechanism and equilibrium}

\begin{figure}[ht]
\centering
\includegraphics[width=8.0cm, height=8.0cm ]{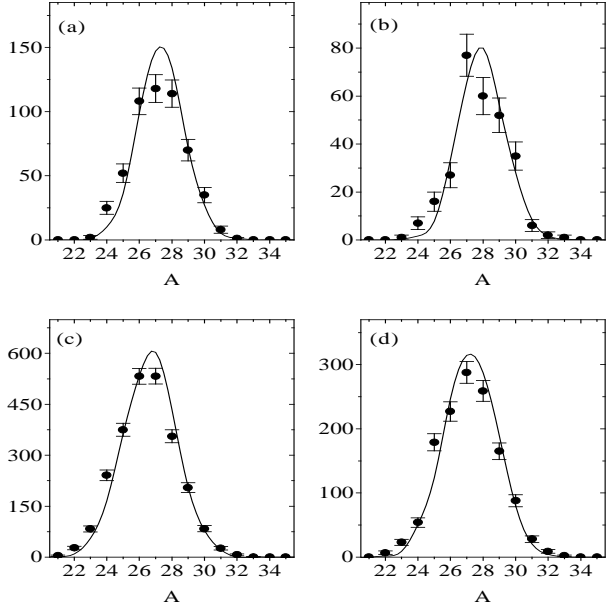}

\caption{ Experimental ( solid circles ) and
simulated ( solid lines ) mass distributions
for the fully isotopically resolved quasiprojectiles with $ Z_{tot}=14 $ 
\cite{SiSnNExch},
(a) - $^{28}$Si(30AMeV) + $^{112}$Sn,
(b) - $^{28}$Si(30AMeV) + $^{124}$Sn,
(c) - $^{28}$Si(50AMeV) + $^{112}$Sn,
(d) - $^{28}$Si(50AMeV) + $^{124}$Sn.
}
\label{fgnzqp}
\end{figure}

In order to establish whether the fragment data from reactions 
$^{28}$Si+$^{124,112}$Sn at projectile energy 50 and 30 AMeV 
\cite{SiSnNExch} 
originate from statistical decay of hot quasiprojectiles, it is necessary to 
investigate the dynamical properties of the projectile-like source 
and to determine the production mechanism. The observed fragment 
data \cite{RLSiSn,SiSnNExch} provide full information 
( with the exception of emitted 
neutrons ) on the decay of thermally equilibrated hot quasi-projectiles 
with known mass ( A=20-30 ), charge, velocity and excitation energy. 
Detailed investigation of the reaction mechanism \cite{SiSnNExch} 
allowed to establish a dominant reaction scenario. 
Excellent description of the fragment observables 
was obtained using the model of deep-inelastic transfer ( DIT ) \cite{DITTGSt} 
for the early stage of collisions and the statistical multifragmentation 
model ( SMM ) \cite{SMM} for de-excitation. The model describes well the 
dynamical properties of the reconstructed quasi-projectile such as center 
of mass velocity, excitation energy ( see Fig. \ref{exsim} ) 
and isospin-asymmetry ( see Fig. \ref{fgnzqp} ). The fragment observables 
such as multiplicity, charge distributions and mean values of N/Z for 
a given charge were also reproduced reasonably well \cite{SiSnNExch}. 
Thus the data 
can be considered as well understood in terms of reaction mechanism. 

The observation that a wide range of observables is in agreement with 
the dominant reaction scenario of the deep inelastic transfer 
( binary collision with intense nucleon exchange between the projectile 
and target nuclei, each of them represented by a statistically equilibrated 
Fermi gas ) followed by statistical multifragmentation 
( assuming statistically 
equilibrated freeze-out configuration ) allows to conclude that 
the assumption of statistical equilibrium is applicable and that 
thermodynamical observables can be defined for the hot quasi-projectile. 
The experimentally observed exponential scaling behavior shown 
in Figs. \ref{fgy3rat} and \ref{Iso50} further demonstrates compliance with 
grand-canonical picture. The contribution from non-equilibrium processes 
such as pre-equilibrium emission was shown to be weak \cite{SiSnNExch}. 
The number 
of emitted neutrons, which are not detected, is, according 
to simulation, between one and two per event and thus no significant 
distortion of the picture can be expected. The observed quasi-projectile 
multifragmentation data can thus be used for investigation 
of thermodynamical properties of the multifragmentation source and 
of the possible phase transition.

\subsection*{Thermodynamical observables and isospin-asymmetric 
phase transition}

In order to investigate thermodynamical properties of the 
hot multifragmentation source, one needs to characterize a set 
of observables, including extensive ones such as thermal excitation 
energy ( heat ) and intensive ones such as temperature and chemical 
potentials. 

\begin{figure}[ht]
\centering
\includegraphics[width=8.0cm, height=7.0cm ]{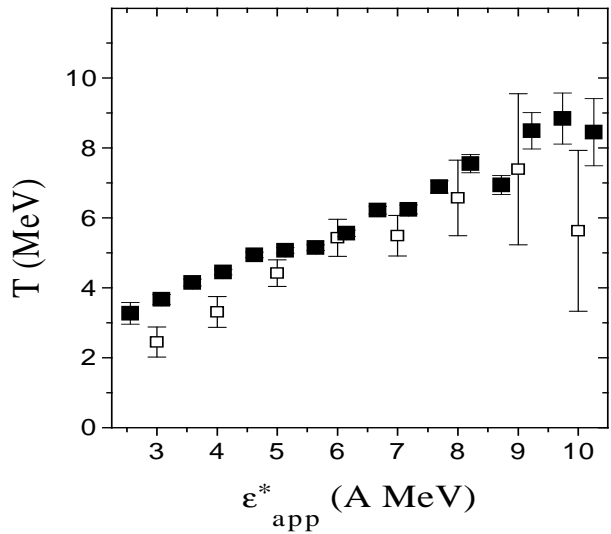}
\caption{
Dependence of the temperature $ T $ on excitation energy $\epsilon^{*}_{app}$. 
Solid squares - temperature determined 
using the double isotope ratio thermometer  \hbox{d,t/$^{3}$He,$^{4}$He}. 
Open squares - temperature determined from the dependence of the yield ratio
Y($^{3}$H)/Y($^{3}$He) on the N/Z ratio of the quasiprojectile 
\protect\cite{MuTemp}. }
\label{sisntemp}
\end{figure}

For the quasi-projectile data from the reactions 
$^{28}$Si+$^{124,112}$Sn at projectile energy 30 and 50 AMeV, 
temperature was determined both using double isotope ratio method and the 
N/Z-dependence of the isobaric yield ratio Y($^{3}$H)/Y($^{3}$He) \cite{MuTemp}. 
The results are shown in Fig. \ref{sisntemp}. Both methods seem to 
provide consistent results. Double isotope ratio temperature 
exhibits a short plateau starting around $\epsilon^{*}_{app}$=4.5 AMeV,  
while the Y($^{3}$H)/Y($^{3}$He) temperature seems to exhibit similar 
feature at somewhat higher excitation energy. The presence of a 
plateau in the caloric curve was interpreted in the literature \cite{Pocho} 
as a signature of the 
first-order phase transition, nevertheless the influence of additional 
effects such as non-thermal flow or variation of the source size 
may affect the shape of the observed caloric curve \cite{JBNCalCurv} 
and such a signal can not 
be considered as an unambiguous signature without providing further 
evidence. For the data in Fig. \ref{sisntemp} 
the influence of non-thermal flow and variation of the source size 
can be excluded due to well understood reaction mechanism. 

    \begin{figure}[ht]                                        
\centering
\includegraphics[width=8.0cm, height=8.0cm ]{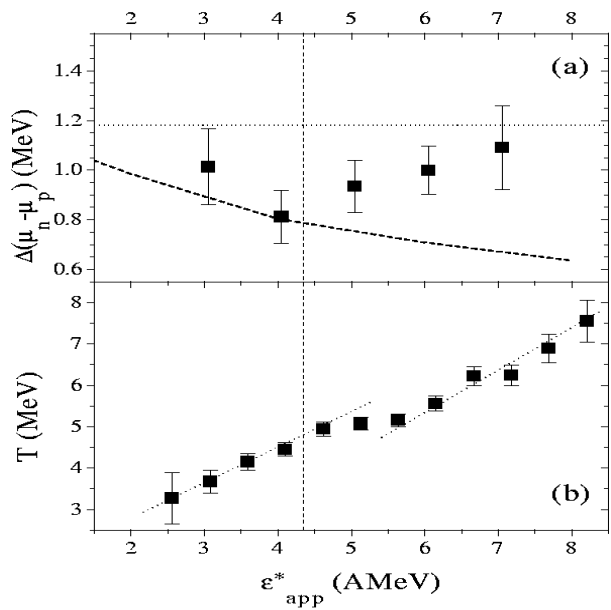}

    \caption{
    (a) The dependence of observable $\beta^{\prime} T$ 
    ( $=\Delta(\mu_n-\mu_p)$ ) on relative 
    excitation energy of the quasiprojectile \cite{SiSnIso} ( full squares ). 
    Horizontal line shows estimate of $\beta^{\prime} T$ obtained when 
    assuming $\mu_{p,n} \approx -S_{p,n}$.
    Dashed curve represents the expected trend for homogeneous system.
    (b) Corresponding caloric curve \cite{MuTemp}. 
    Vertical line indicates the position of turning point. }
    \label{IsoTmp}
    \end{figure}

Of the two methods for temperature determination, the Y($^{3}$H)/Y($^{3}$He) 
temperature relies on additional assumption concerning chemical potentials. 
The insight into behavior of the chemical potentials can be obtained 
using the isoscaling analysis. The values of the isoscaling parameter 
$\beta^{\prime}$ for the data in Fig. \ref{Iso50} can be related, according 
to formula (\ref{r21isobt}), to the isovector component 
of the chemical potential. In Fig. \ref{IsoTmp}a 
we present an estimate of the difference of isovector chemical potential 
using the observable $\beta^{\prime} T$, canceling
out a trivial 1/T-dependence of the isoscaling
parameter \cite{SiSnIso}. The double isotope ratio 
temperature \cite{Albergo} from the isotopic ratios
Y($^{2}$H)/Y($^{3}$H) and Y($^{3}$He)/Y($^{4}$He), 
which is independent of assumptions concerning chemical potentials, 
was used. The horizontal line represents 
an estimate of the zero temperature value
of $\Delta(\mu_n-\mu_p)$ by the proton and neutron
separation energies ( $\mu_{p,n} \approx -S_{p,n}$ )
of reconstructed quasi-projectiles, using the known N/Z equilibration 
and the correction for neutron emission 
from back-tracing of DIT/SMM simulations \cite{SiSnNExch}. 
The zero temperature estimates of $\Delta(\mu_n-\mu_p)$ for subsets of data in 
all excitation energy bins are consistent with the horizontal line. 
Any experimental deviations 
from the horizontal line can be understood as
a non-trivial dependence representing the details of de-excitation.
The experimental dependence of $\beta^{\prime} T$ 
indicates initial decrease at low excitation energies
( consistent with expansion of the homogeneous hot source ), 
further it exhibits a turning-point at 4 AMeV
followed by increase toward the zero temperature value at 6-7 AMeV.
The estimate of $\beta^{\prime} T$ for homogeneous system ( dashed curve ) 
was obtained assuming the Fermi-gas $\rho^{2/3}$-dependence and using 
the estimate of free 
volume by formula (\ref{eqnchi}) for each excitation energy bin. 
The trend explains a reasonable success of the $\mu_{p,n} \approx -S_{p,n}$ 
approximation used to extract the mirror nucleus temperature in 
Fig. \ref{sisntemp}, where the viability of the approximation 
at high temperatures
was supported by a possible counterbalance of two effects, namely expansion
of the hot nucleus ( leading to a decrease of the absolute values 
of chemical potentials ) and chemical separation into an isospin
symmetric heavy fraction and an isospin asymmetric nucleon gas, leading
to increase of $\mu_{n} - \mu_{p}$ of the dilute phase ( nucleon gas ). 
Since the isoscaling parameter depends directly on the free nucleon chemical 
potentials,
the turning-point at 4 AMeV can be understood as a signal of the onset
of chemical separation which reverts the decrease of the free nucleon
chemical potential consistent with expansion of the homogeneous system.
As demonstrated by comparison with the caloric curve in Fig. \ref{IsoTmp}b  
( obtained using double isotope ratio thermometer without assumption 
concerning chemical potentials ) 
the onset of chemical separation is correlated to the onset 
of the plateau in the caloric curve, thus signaling that chemical 
separation is accompanied by latent heat. Such a behavior can be 
expected in the first-order phase transition. The dotted lines 
in Fig. \ref{IsoTmp}b indicate the discontinuity in the heat capacity 
of the system. 

\begin{figure}[htbp]
\centering
\includegraphics[width=9.0cm, height=6.0cm ]{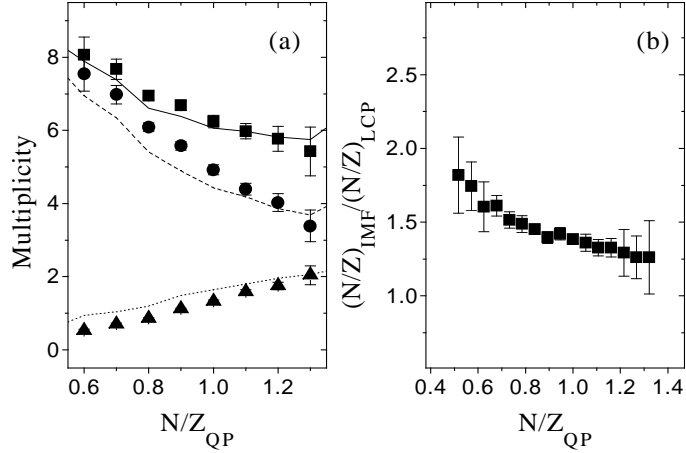}
\caption{
(a) - Multiplicity of charged fragments (squares), LCPs (circles)
and IMFs (triangles) versus
N/Z$_{QP}$. Corresponding lines represent the DIT/SMM calculation.
(b) - Experimental ratio of the mean values of N/Z of LCPs
and IMFs versus the N/Z$_{QP}$.
Data are given for the reaction $^{28}$Si(50AMeV)+$^{112}$Sn 
\cite{IsoDist}.
}
\label{mcpnz}
\end{figure}

\begin{figure}[htbp]
\centering
\includegraphics[width=8.0cm, height=6.0cm ]{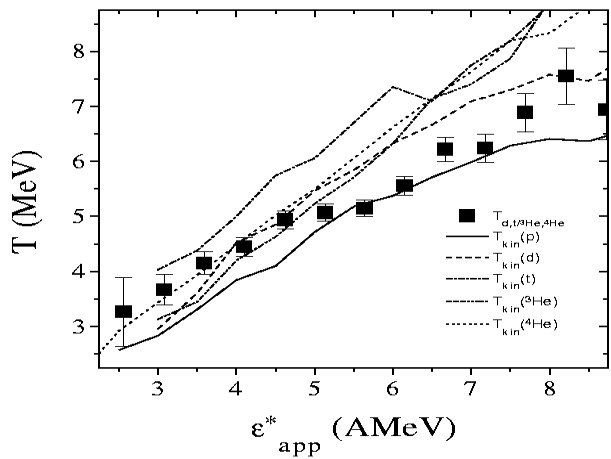}
\caption{
Comparison of temperature obtained using the double isotope ratio thermometer 
\hbox{d,t/$^{3}$He,$^{4}$He} \cite{MuTemp} with kinematic temperatures 
of p,d,t,$^{3}$He,$^{4}$He in the c.m. frame of the quasi-projectiles. 
}
\label{fgslptmp}
\end{figure}

The behavior, observed in Fig. \ref{IsoTmp}, is consistent with the 
first-order phase transition. Several signals of the first-order 
phase transition have been reported in the recent experimental studies, 
such as negative micro-canonical heat capacity \cite{DAgo}, characterized 
by abnormal kinetic energy fluctuation,  and the "fossil" signal 
of spinodal decomposition \cite{Borderie}, characterized by abnormal production 
of equally sized fragments. These indications of the  first-order 
phase transition are based on careful event-by-event analysis of 
the multifragmentation data combined with model assumptions 
and results of theoretical simulations. The isospin-asymmetry 
of the fragments is not explored in such analyzes. The behavior 
in Fig. \ref{IsoTmp} relates the signal of chemical separation 
to the plateau of the caloric curve, thus providing a signal 
of isospin-asymmetric 
first-order phase transition. Further support for such conclusion 
can be deduced from Fig. \ref{mcpnz}b, demonstrating the inhomogeneous 
distribution of isospin-asymmetry \cite{IsoDist} among dense and dilute phases 
represented by light charged particles and intermediate mass 
fragments, respectively. The effect is strongly dependent 
on the N/Z ratio of the quasiprojectile. So are multiplicities 
of LCPs and IMFs in Fig. \ref{mcpnz}a which further indicate that  
with increasing isospin-asymmetry of the quasiprojectile the dilute 
phase becomes more isospin-asymmetric and abundant. 
The observations related to Figs. \ref{IsoTmp}, \ref{mcpnz} are 
consistent with theoretical predictions of the isospin-asymmetric 
phase transition in nuclear media. The estimated density 
of the homogeneous system at the turning point in Fig. \ref{IsoTmp}a 
is consistent with theoretical estimates of the position and shape
of the spinodal contour, typically at almost constant 
total density 0.6$\rho_0$ in a wide range of asymmetries \cite{Spinod}. 
Thus, inside the spinodal region the homogeneous nuclear medium 
is quickly replaced by both isospin and spatially inhomogeneous system. 
Such a conclusion is further supported by a comparison 
of the temperature obtained using the double isotope ratio thermometer 
\hbox{d,t/$^{3}$He,$^{4}$He} \cite{MuTemp} with kinematic temperatures 
of p,d,t,$^{3}$He,$^{4}$He obtained from Maxwellian fits of the particle 
kinetic energy spectra in the quasiprojectile frame 
( see Fig. \ref{fgslptmp} ). While at low excitation energies  
the double isotope ratio thermometer \hbox{d,t/$^{3}$He,$^{4}$He} 
represents essentially an average of the kinematic temperatures of 
individual species, in the plateau region occurs a 
transition to the regime where the double isotope ratio temperature 
represents the kinematic temperature of protons ( and possibly of the nucleon 
gas ). Such a transition, just above the turning-point where 
the isovector part of chemical potential deviates from the trend 
of homogeneous system (as shown in Fig. \ref{IsoTmp}), suggests 
that the isotopic composition of fragments is 
in equilibrium with the free nucleon gas what is a basic assumption 
used in grand-canonical models such as SMM \cite{SMM}. The higher values of 
kinematic temperatures of d,t,$^{3}$He,$^{4}$He suggest that the clusters are 
created prior to such equilibration ( possibly by spinodal decomposition ). 
On the other hand, the low 
energy behavior can be explained by an assumption that the isotopic 
composition is determined at the instant when clusters are emitted 
what is in good agreement with our understanding of de-excitation 
of the compound nucleus \cite{HausFesch}. The two different values of 
heat capacities in low- and high-energy parts of the caloric curve, 
suggested in Fig. \ref{IsoTmp}b, can thus be understood as 
representing a homogeneous nuclear liquid and an isospin-asymmetric 
free nucleon gas, respectively.

Apart from signatures of the first-order phase transition, several 
experimental works \cite{MFCrit,SBDFail}, exploring observed mass ( charge ) 
distributions, report a critical behavior in the 
multifragmentation data, which can be understood as a signal of the 
second-order phase transition at the critical temperatures between 
5 - 10 MeV \cite{CritTemp}. However, multifragmentation studies employing the 
statistical model analysis of charge distribution \cite{Karnau} 
suggest the value of critical temperature above 15 MeV. 
Besides of raising obvious questions concerning the applicability of used 
methods of analysis, such contradictory observations may in principle 
suggest a complex phase diagram, possibly analogous to the Ising model, 
where both first-order and second-order can be observed. It is also possible 
that the observed critical behavior is a specific feature of the finite 
system, as suggested in theory \cite{FinPhTr}. 
The recent experimental work where the critical behavior is related to 
the size of largest fragment as an order parameter \cite{FinExp} supports such 
assumptions. The finite-size effect can play role also in the 
multifragmentation of quasiprojectiles with A=20-30 and thus similar 
analysis for heavier sources is necessary to evaluate the role of such effect 
and establish the link to the phase transition in the isospin-asymmetric 
nuclear matter at sub-saturation densities. It may be however noted that 
theoretical studies of small systems \cite{FinThDyn} suggest that 
van der Waals-like 
loops observed in infinite systems are affected in transition to finite system 
only by small terms and thus the conclusions for finite systems can be 
in principle applicable also for infinite case. 

\section*{Summary and conclusions}

In this article we presented an overview of the recent progress 
in the studies of nuclear multifragmentation. 
Theoretical and experimental concepts relevant to investigation of 
isotopic trends in nuclear multifragmentation were described 
and the possibilities to extract the physical information
on the nuclear equation of state were outlined. 
A special emphasis was put on the methods how 
the isotopic composition of fragments can be used to extract the values 
of thermodynamical observables of the system
undergoing multifragmentation such as temperature and 
chemical potentials ( free nucleon densities ). 
Various methods for extraction of thermodynamical variables were presented and 
possibilities of their use were documented on examples of multifragmentation 
data. It was shown that the yield ratios of specific isotopes from statistical 
multifragmentation typically exhibit an exponential scaling with various 
observables. The logarithmic slopes of such dependences can be 
related to thermodynamical observables such as grand-canonical 
temperatures ( isotopic thermometers ) or chemical potentials ( isoscaling ). 
Finally, an extensive overview of investigations of multifragmentation 
of the fully reconstructed quasi-projectiles is presented. The dominant 
reaction mechanism is determined as binary nucleus-nucleus collisions 
with intense dissipation and heating via nucleon exchange. The 
implications of the fragment data concerning the isospin-asymmetric 
liquid-gas phase transition in the nuclear matter are discussed. 
Multiple correlated signals point toward observation of a first-order 
phase-transition in the quasi-projectile multifragmentation data. 

The author acknowledges fruitful collaboration with research groups 
of Professors S.J. Yennello and J.B. Natowitz during his stay 
at Cyclotron Institute of Texas A\&M University. Examples from the  
analysis of experimental data acquired using FAUST and NIMROD arrays 
are used throughout this work to illustrate the use of reviewed methods. 
Insightful comments from A.I. Sanzhur and P.I. Zarubin are gratefully 
acknowledged. This work was supported by the Slovak Scientific Grant Agency 
through grant VEGA-2/1132/21.


\begin{thebibliography}{00}

\bibitem{ExpMfrag}
L.G. Moretto, G.J. Wozniak // Ann. Rev. Nucl. Part. Sci. 1993. V.43. P.379. 

\bibitem{ExpMfragRec}
Proc. of Int. Conf. CRIS-2000, Catania, 2000. Nucl. Phys. A. 2001. V.681;  
Proc. of Int. Conf. NN-2003, Moscow, 2003. Nucl. Phys. A. 2004. V.734. 

\bibitem{AstroSpall}
R. Silberberg, C.H. Tsao // Phys. Rep. 1990. V.191. P.351.

\bibitem{Richert}
J. Richert, P. Wagner // Phys. Rep. 2001. V.350. P.1.

\bibitem{SMM}
J.P. Bondorf et al. // Phys. Rep. 1995. V.257. P.133.

\bibitem{MullerSerot}
H. M\"uller, B.D. Serot // Phys. Rev. C. 1995. V.52. P.2072.

\bibitem{Spinod}

B.-A. Li, C. Ko // Nucl. Phys. A. 1997. V.618. P.498; 
V. Baran et~al. // Phys. Rev. Lett. 2001. V.86. P.4492; 
M. Colonna et~al. // Phys. Rev. Lett. 2002. V.88. P.122701; 
J. Marqueron, P. Chomaz // Phys. Rev. C. 2003. V.67. P.41602(R).

\bibitem{MMMC}
D.H.E. Gross // Phys. Rep. 1997. V.279. P.119.

\bibitem{FinThDyn}
Terrell L. Hill // Thermodynamics of small systems, W.A. Benjamin Publishers, 
New York, Amsterdam, 1963. 


\bibitem{SiSnNExch}
M. Veselsky et al. // Phys. Rev. C. 2000. V.62. P.064613.

\bibitem{SnAlMARS}
M. Veselsky et al. // Nucl. Phys. A. 2003. V.724. P.431.

\bibitem{HausFesch}
W. Hauser, H. Feshbach // Phys. Rev. 1952. V.87.  P.366.

\bibitem{AsymFiss}
L.G. Moretto // Nucl. Phys. A. 1975. V.247. P.211.

\bibitem{McMahan}
M.A. McMahan et al. // Phys. Rev. Lett. 1985. V.54 P.1995.

\bibitem{GEMINI}
R. Charity et al. // Nucl. Phys. A. 1988. V.483. P.391.

\bibitem{MFrgTrans}
J.A. Lopez, J. Randrup // Nucl. Phys. 1989. A. V.503. P.183.

\bibitem{SBDFail}
P. Kreutz et al. // Nucl. Phys. A. 1993. V.556. P.672.

\bibitem{TdepHF}
G. Sauer, H. Chandra, U. Mosel // Nucl. Phys. A. 1976. V.264. P.221.

\bibitem{DropBubbl}
P.J. Siemens // Nature. 1983. V.305. P.410.

\bibitem{KolSanzh}
V.M. Kolomietz et al. // Phys. Rev. C. V.64. 2001. P.24315.

\bibitem{Fisher}
M.E. Fisher // Physics. 1967. V.3. P.255.




\bibitem{SMMOrder}
K.A. Bugaev et al. // Phys. Rev. C. 2000. V.62. P.044320; 
Phys. Lett. B. 2001. V.498. P.144; 
P.T. Reuter, K.A. Bugaev // Phys. Lett. B. 2001. V.517. P.233.

\bibitem{Ising}
J.M. Hammersley // Proc. of 87th Intern. Colloq. CNRS, Paris, 1957. P.17.


\bibitem{Prod}
M. Veselsky // Nucl. Phys. A. 2002. V.705. P.193.

\bibitem{NSkin}
G.A. Souliotis et al. // Phys. Lett. B. 2002. V.543. P.163; 
Phys. Rev. Lett. 2003. V.91. P.022701.

\bibitem{INC}
M. P. Guthrie, R. G. Alsmiller, H. W. Bertini // 
Nucl. Instrum. Meth. 1968. V.66. P.29;  
V.S. Barashenkov, V.D. Toneev // 
High Energy interactions of particles and nuclei with nuclei 
(in russian), 1972.

\bibitem{SpecMFrg}
A. Schuttauf et al. // Nucl.Phys. A. 1996. V.607. P.457.

\bibitem{TDHF}
J.W. Negele // Rev. Mod. Phys. 1992. V.54. P.913.

\bibitem{VUU}
H. Kruse, B.V. Jacak, H. Stoecker // Phys. Rev. Lett. 1985. V.54. P.289;
J. Molitoris, H. Stoecker // Phys. Rev. C. 1985. V.32. P.346;
J. Molitoris, H. Stoecker, B. Winer // Phys. Rev. C. 1987. V.36. P.220.

\bibitem{BUU}
G. Bertsch, S. Gupta // Phys. Rep. 1988. V.160. P.189; 
W. Cassing et al. // Phys. Rep. 1990. V.188. P.363.

\bibitem{LV}
C. Gregoire et al // Nucl.Phys. A. 1987. V.465. P.317; 
Nucl.Phys. A. 1987. V.471. P.399c;
P. Schuck et al. // Prog. Part. Nucl. Phys. 1990. V.22. P.181.

\bibitem{BNV}
A. Bonasera et al. // Phys. Rep. 1994. V.243. P.1.

\bibitem{QMD}
J. Aichelin // Phys. Rep. 1991. V.202. P.233.

\bibitem{FAUST}
F. Gimeno-Nogues  et al. // Nucl. Instrum. and Meth. A. 1997. V.399. P.94.

\bibitem{NIMROD}
N. Marie et al. // Progress in Research, 1997--1998, Cyclotron Institute,
Texas A\&M University, College Station, 1998. P.V-19; 
R. Wada et al. // Progress in Research, 1998--1999, 
Cyclotron Institute, Texas A\&M University, College Station, 1999. P.V-15.

\bibitem{MVCal}
M. Veselsky et al. // Progress in Research, 2000--2001, Cyclotron Institute, 
Texas A\&M University, College Station, 2001. P.V-13.

\bibitem{TGCal}
L. Tassan-Got // Nucl. Instrum. Meth. B. 2002. V.194. P.503.

\bibitem{DANF2001}
M. Veselsky, G.A. Souliotis, S.J. Yennello // Proc. 5th Int. Conf. DANF-2001, 
Casta-Papiernicka, Slovakia, 2001. World Scientific, 2002. P.461.

\bibitem{Shetty}
D.V. Shetty et al. // Phys. Rev. C. 2003. V.68. P.54605.

\bibitem{Frank}
J.D. Frankland et al. // Nucl. Phys. A. 2001. V.689. P.905.

\bibitem{Cavata}
C. Cavata et al. // Phys. Rev. C. 1990. V.42. P.1760.

\bibitem{YRat}
R. Wada et al. // Phys. Rev. Lett. 1987. V.58. P.1829; 
H.W. Barz et al. // Phys. Lett. B. 1988. V.211. P.10; 
S.J. Yennello et al. // Phys. Lett. B. 1994. V.321. P.14; 
H. Johnston et al. // Phys. Lett. B. 1996. V.371. P.186; 
E. Ramakrishnan et al. // Phys. Rev. C. 1998. V.57. P.1803.

\bibitem{ColdBkup}
H. Fuchs, K. M\"ohring // Rep. Prog. Phys. 1994. V.57. P.231.

\bibitem{MuTemp}
M. Veselsky et al. // Phys. Lett. B. 2001. V.497. P.1.

\bibitem{Albergo}
S. Albergo et al. // Nuovo Cimento 1995. V.89. P.1.

\bibitem{KoonRand}
J. Randrup, S.E. Koonin // Nucl. Phys. A. 1981. V.356. P.223.

\bibitem{AlbExp}
J.A. Hauger et al. // Phys. Rev. Lett. 1995. V.77. P.235; 
H. Xi et al. // Z. Phys. A. 1997. V.359. P.397; 
A. Ruangma et al. // Phys. Rev. C. 2002. V.66. P.44603.

\bibitem{AlbCorr}
M.B. Tsang et al. // Phys. Rev. Lett. 1997. V.78. P.3836.

\bibitem{IsoDist}
M. Veselsky et al. // Phys. Rev. C. 2000. V.62. P.41605(R).

\bibitem{Xu}
H.S. Xu et al. // Phys. Rev. Lett. 2000. V.85. P.716.

\bibitem{TsangIso}
M.B. Tsang et al. // Phys. Rev. Lett. 2001. V.86. P.5023.

\bibitem{BotvIso}
A.S. Botvina et al. // Phys. Rev. C. 2002. V.65. P.44610.

\bibitem{Lozhkin}
O.V. Lozhkin, W. Trautmann // Phys. Rev. C. 1992. V.46. P.1996.

\bibitem{GSHRIso}
G.A. Souliotis et al. // Phys. Rev. C. 2003. V.68. P.24605.

\bibitem{Fisiso}
M. Veselsky, G.A. Souliotis, M. Jandel //  Phys. Rev. C. 2004. V.69. P.44607.

\bibitem{GSNZEq}
G.A. Souliotis et al. // Phys. Lett. B. 2004. V.588. P.35.

\bibitem{SiSnIso}
M. Veselsky, G.A. Souliotis, S.J. Yennello // 
Phys. Rev. C. 2004. V.69. P.31603(R).

\bibitem{RLSiSn}
R. Laforest et al. // Phys. Rev. C. 1999. V.59. P.2567.

\bibitem{DITTGSt}
L. Tassan-Got, C. St\'{e}fan // Nucl. Phys. A. 1991. V.524. P.121.

\bibitem{Pocho}
J. Pochodzalla et al. // Phys. Rev. Lett. 1995. V.75. P.1040.

\bibitem{JBNCalCurv}
J.B. Natowitz et al. // Phys. Rev. C. 2002. V.65. P.34618.

\bibitem{DAgo}
M. D'Agostino et al. // Nucl. Phys. A. 1999. V.650. P.329.

\bibitem{Borderie}
B. Borderie et al. // Phys. Rev. Lett. 2001. V.86. P.3252.

\bibitem{MFCrit}
R.W. Minich et al. // Phys. Lett. B. 1982. V.118. P.458; 
A.D. Panagiotou et al. // Phys. Rev. Lett. 1984. V.52. P.496; 
J.B. Elliot et al. // Phys. Rev. C. 2000. V.62. P.64603.

\bibitem{CritTemp}
J.B. Elliot et al. // Phys. Rev. Lett. 2002. V.88. P.42701; 
M. Kleine Berkenbusch et al. // Phys. Rev. Lett. 2002. V.88. P.22701.

\bibitem{Karnau}
V.A. Karnaukhov et al. // Phys. Rev. C. 2003. V.67. P.011601(R).

\bibitem{FinPhTr}
T.S. Biro, J. Knoll, J. Richert // Nucl. Phys. A. 1986. V.459. P.692; 
R. Botet, M. Ploszajczak // Phys. Rev. E. 2000. V.62. P.1825.

\bibitem{FinExp}
J.D. Frankland et al. // arXiv.org: e-print nucl-ex/0404024, 2004.

\end{thebibliography}
\end{document}